\documentclass[final,5p,times,twocolumn]{elsarticle}

\usepackage{geometry} % see geometry.pdf on how to lay out the page. There's lots.
\usepackage{booktabs} % 专业表格格式
\usepackage{multirow} % 支持多行合并
\usepackage{adjustbox} % 推荐使用此宏包
\usepackage{graphicx} % 确保已导入graphicx
\usepackage{siunitx} % 数值对齐
\usepackage{makecell}    % 多行表头支持
\usepackage{enumitem}      % 用于定义环境
\usepackage{bm}           % 用于加粗数学符号
\usepackage[colorlinks=true, allcolors=blue]{hyperref}
\usepackage{array}    % 用于调整列格式
\usepackage{caption}  % 用于定制图表标题
\usepackage{threeparttable} % 用于添加表格注释
\usepackage{float}      % 增强浮动体控制
\usepackage[utf8]{inputenc}
\DeclareUnicodeCharacter{200B}{} % 关键：忽略零宽空格
\newtheorem{definition}{Definition}

\newtheorem{property}{Property}

\usepackage{algpseudocode}
\usepackage{algorithm}

\newcommand{\R}{\mathbb{R}}
\usepackage{amssymb}
\usepackage{amsmath}
\journal{Knowledge-Based Systems}

\begin{document}

\begin{frontmatter}



\title{Multi-view Attention Fusion of Heterogeneous Hypergraph with Dynamic Behavioral Profiling for Personalized Learning Resource Recommendation}
\author{Tao Xie\corref{cor1}\fnref{label1}}
\author{Yan Li\fnref{label1}}
\author{Yongpan Sheng\fnref{label2}}
\author{Jian Liao\fnref{label1}}
\fntext[label1]{Tao Xie, Yan Li, and Jian Liao are with the Faculty of Education, Southwest University, Chongqing, China 400715}

\fntext[label2]{Yongpan Sheng is with the College of Computer and Information Science, Southwest University, Chongqing, China 400715}

\cortext[cor1]{xietao@swu.edu.cn}


%%\author{Tao Xie, Yan Li, Yongpan Sheng, and Jian Liao} %% Author name

%% Abstract
\begin{abstract}
Hypergraph can capture complex and higher-order dependencies among learners and learning resources in personalized educational recommender systems. Many existing hypergraph-based recommendation approaches underexplored the dynamic behavioral processes inherent to learning and often oversimplified the complementary information embedded across multiple dimensions (i.e. views) within hypergraphs. These limitations compromise both the distinctiveness of learned representations and the model's generalization capabilities, especially under data-sparse conditions typical in educational settings. In this study, we propose a unified model comprising a dynamic behavioral profiling module and a multi-view attention fusion module based on heterogeneous hypergraph construction. The dynamic behavioral profiling module is designed to capture evolving behavioral processes and infer latent higher-order relations crucial for hypergraph completion; The multi-view fusion module cohesively integrates information from distinct relational views, enriching the overall data representation. The proposed model was systematically evaluated on five public benchmark datasets and one real-world, self-constructed dataset. Experimental results demonstrate that the model outperforms baseline methods across most datasets in key metrics; Furthermore, hypergraph completion based on dynamic behavioral profiling contributes significantly to performance gains, though its efficacy is modulated by dataset characteristics.  Beyond offline experiments, we implemented a functional prototype system tailored for postgraduate student literature recommendation. A mixed-methods user study was conducted to assess its practical utility. Quantitative analysis revealed significantly higher perceived recommendation quality; Qualitative feedback highlighted enhanced user engagement and satisfaction with the prototype system. This work advances hypergraph-based personalized educational recommender systems and offers human-centered insights for designing adaptive learning environments. 

\end{abstract}



%% Keywords

\end{frontmatter}



\section{Introduction}
\label{sec1}
%% Labels are used to cross-reference an item using \ref command.
The global expansion of online education, combined with fragmented learning challenges, highlights the critical need for effective information technology mediation \cite{van2025factors}. Personalized Educational Recommender Systems (PERS) are used to help learners overcome information overload and knowledge disorientation \cite{montaner2003taxonomy, martinez2025towards}. By effectively matching knowledge levels with appropriate learning resources, PERS can significantly enhance learning motivation and improve educational outcomes \cite{el2023video}. Unlike recommender systems in e-commerce and digital entertainment, the core goal of PERS is not only to match user interests by establishing associations between users and items but also to promote cognitive development \cite{huang2023effects}. Educational research supports that PERS should be able to accurately model interactions between learners and learning resources, semantic relationships among learning resources per se, and dynamic profiling learners based on cognitive states \cite{ain2024learner, li2025breaking}.

PERS has evolved from collaborative filtering to graph neural networks, demonstrating that graphs are an effective tool for modeling complex learner-resource associations in educational information systems \cite{liu2025graph, albreiki2023extracting}. However, traditional graph structures, such as user-item bipartite graphs, social relation-based graphs, and knowledge graphs, only represent binary pairwise relationships, making it difficult to efficiently model the complex, higher-order dependencies prevalent in educational interactions \cite{lin2023automatic}. Hypergraphs provide a more expressive alternative by connecting multiple nodes simultaneously via hyperedges, thereby representing multi-dimensional educational relationships \cite{antelmi2023survey}. Yet, many hypergraph-based recommender systems often create hypergraphs with multiple relationship types, each representing an independent view or information dimension \cite{ma2024cross, liu2025multi}. Although these views capture specific high-order correlations between learners and learning resources, allowing for local descriptions of complex systems, most existing methods struggle to effectively fuse the unique, complementary information inherent to each view. Concatenating feature vectors from different perspectives can obscure unique structural semantics, preventing deep integration of effective information during representation learning \cite{ao2024hypergraph, li2024multi}.

Recent research has focused increasingly on multi-view hypergraph recommendation models. However, their application is still limited to specific domains and has not yet matured in PERS \cite{lai2023multi,zeng2025enhanced,liu2024dual}. This insufficiency can not effectively address the practical requirements of PERS for modeling complex interactions between learners and learning resources, semantic relationships, and learners' dynamic behaviors \cite{ain2024learner, li2025breaking}. Some studies homogenize such interactions by compelling heterogeneous nodes and relations into a unified semantic space without integration of multi-view information, compromising the distinctiveness of representations and the model's generalization capabilities \cite{li2024hje, hayat2024heterogeneous}. Scholars argued that hypergraph projection inherently risks losing critical higher-order structural information, quantifying the combinatorial impossibility of recovering these lost structures without specific interventions \cite{wang2024graphs} and traditional embeddings frequently fail to capture multifaceted non-pairwise dynamics when complex interactions are forcibly reduced to simplified representations \cite{hayat2024heterogeneous}. As a result, the ultimate representations of learners and learning resources often lack the fine-grained, selective cross-view information complementation required to significantly improve the recommender system's performance.

Moreover, the learning process within PERS exhibits continuous evolution driven by learning interactions, new knowledge acquisition, shifting personal interests, and phased cognitive development \cite{wang2025mmkt, cheng2024dygkt, guo2023understanding}. Some research has begun to explore dynamic hypergraph constructions to model temporal changes in recommender systems \cite{wei2022dynamic}. However, many existing hypergraph-based recommendation models, particularly within educational contexts, primarily rely on static or quasi-static, predefined hypergraph structures that lack dynamic adaptability to capture evolving learning processes and behavioral changes \cite{yang2022multi, guo2024multi}. These models typically create a single global hypergraph from all historical interactions, segment temporal data into fixed time windows for snapshot hypergraphs \cite{10.1145/3653306}, or incorporate static side information during initial hypergraph construction \cite{ao2024hypergraph, 10.1145/3711896.3736885}. This reliance on static modeling results in outdated and biased user representations, causing predictions to deviate further from users' true intentions and cognitive needs. Consequently, not only is real-time recommendation accuracy diminished, but the core educational function of providing personalized learning resources tailored to update cognitive states is also compromised.

To address the aforementioned limitations, this paper proposes a novel framework called Multi-view Attention Fusion with Dynamic Behavioral Profiling of Heterogeneous Hypergraphs. To begin, we create a heterogeneous hypergraph by combining a user-item bipartite graph with item-associated semantic information to model rich, higher-order semantic relations. Second, to capture the dynamic nature of user-item interactions, we create behavioral profiles for users. These profiles are then used to infer new hyperedges, which improves the original hypergraph's structural information and dynamic adaptability. To fully capitalize on the structural richness of this heterogeneous hypergraph, we use hypergraph random walk sampling to generate multiple views, each designed to capture unique and complementary information. We then create a multi-view attention fusion module. This module not only learns robust node embeddings within each view, but it also adaptively integrates multi-view information at the representation level using a learnable attention mechanism. It directly addresses the previously mentioned challenge of ineffective information fusion and produces comprehensive representations of learners and learning resources. The main contributions of this work are summarized as follows:

\begin{itemize}
    \item We propose a heterogeneous hypergraph completion framework based on dynamic user behavior that augments the hypergraph's structure and advances the practical use of student profiling for relational learning within PERS.
    \item We present a multi-view attention fusion framework for heterogeneous hypergraphs, significantly enhancing the expressive power of the recommendation model. We innovatively offer a principled methodology to improve model robustness and information comprehensiveness.
    \item We run extensive computational experiments across multiple educational datasets, including five public benchmarks and one self-constructed dataset. These experiments not only confirm the efficacy of the proposed framework, but also demonstrate the generalizability of hypergraph-based recommendation techniques.
    \item Beyond laboratory simulations, we create and test a prototype system for real-world educational settings. This system fills the gap between theoretical hypergraph models and practical applicability by providing validated insights into human-centered educational system design in adaptive learning environments.
\end{itemize}

The remainder of this paper is organized as follows. Section \ref{rw} includes a review of related work. Section \ref{pl} presents essential formal definitions. Section \ref{pm} provides the proposed heterogeneous hypergraph fusion framework. Section \ref{ep} and Section \ref{er} elaborate on the experimental setup and results, respectively. Section \ref{ps} describes the design of the prototype system and reports the results of the user evaluation study. Section \ref{cp} concludes the paper.

\section{Related work}
\label{rw}
\subsection{Personalized educational recommender systems}
Over the past few decades, PERS have been widely adopted in online learning platforms, higher education systems, and even lifelong learning and labor-market-driven education initiatives \cite{amin2024adaptable,pal2025aggregated,li2024creating}. PERS have been used in MOOC platforms to recommend suitable online courses, personalized learning sequences, and cognitively appropriate peers, thereby enhancing student engagement and learning effectiveness \cite{khalid2022literature,tzeng2024personal,elghomary2022design}. In higher education, PERS have been used to create personalized teaching systems that better extract and utilize learning resources \cite{li2024creating}. Pedagogically, PERS can be viewed as information filtering systems that dynamically select appropriate learning resources based on individual differences, such as prior knowledge, learning styles, and cognitive states, to reduce information overload and knowledge disorientation \cite{da2023systematic}. From a technical perspective, alternatively, PERS are viewed as a process of precise decision optimization within complex educational data spaces, centered on designing a matching function that achieves dynamic perception of learning states, in-depth mining of learner–resource relationships, and automated producing learning resource recommendation \cite{zheng2022survey}. Broadly speaking, PERS can be categorized into traditional similarity-based approaches, knowledge graph-enhanced approaches, and context-aware explainable approaches.

Similarity-based approaches successfully applied content-based filtering, collaborative filtering, and hybrid methods to educational settings,  establishing an analytical framework based on user–item interaction matrices \cite{klavsnja2015recommender,amin2023developing}. These methods are relatively simple and computationally efficient. However, they rely heavily on observable behavioral data, e.g., ratings and clicks, while ignoring important unobservable factors in educational activities, such as changes in cognitive states and knowledge structures. Consequently, they often fail to identify the next knowledge point that learners truly require from a knowledge-structure perspective \cite{liu2022recommendation,huang2023effects}. To address this issue, researchers used knowledge graphs as explicit domain knowledge models to improve recommendation accuracy. These methods aim to structurally represent domain knowledge, allowing PERS to perform reasoning based on semantic associations rather than simply mining behavioral co-occurrence patterns \cite{liu2022personalized}. When a learner is interested in a specific knowledge point, PERS can recommend related learning paths or supplementary materials according to the prerequisite or subsequent relationships defined in the knowledge graph, thereby helping learners build a more complete knowledge system \cite{meng2025design}. Knowledge graph-enhanced recommendation partially alleviates data sparsity and cold-start problems through knowledge associations. However, knowledge graphs often simplify complex pedagogical relationships, such as many-to-many cognitive dependencies and dynamic interactive behaviors, into static binary relations \cite{wan2019hybrid}. They do not adequately model learners’ dynamic, multi-dimensional cognitive characteristics, limiting their ability to handle interest drift and temporal variations in cognitive development.

To capture dynamic and interrelated cognitive state representations, researchers used learner profiling to shift recommendations away from static knowledge structures toward prediction based on individual cognitive states. A variety of knowledge tracing and cognitive diagnostic techniques have been employed to recommend learning resources that can address specific attribute deficiencies \cite{abdelrahman2023knowledge,sun2025daskt}. Such methods enhance the explainability and pedagogical relevance of recommendations, but their accuracy heavily depends on predefined, high-quality item–attribute association matrices, i.e., Q-matrices, making them difficult to generalize to open educational scenarios with diverse learner-resource interactions \cite{yun2024doubly}. Recent research has focused on obtaining learner profiles and learning resource representations in unified deep networks, bridging the gap between knowledge graph-based and dynamic learner profile-based approaches. Some scholars, for example, constructed a dual knowledge graph convolutional network for learning resource–knowledge and learner–knowledge domains \cite{dong2024multi} and other scholars employed multi-scale deep reinforcement learning and an attention-based model to characterize learners’ multiple preferences for a multi-scale reinforced profile in personalized recommendation \cite{lin2023multi}. These approaches typically design an evolvable student state representation module that integrates multi-modal temporal data, e.g., answer sequences, viewing durations, and interaction frequencies and outputs a dynamically updated learner embedding, which then interacts with learning resource embeddings for ultimate task prediction \cite{mavromatis2022tempoqr}. 

Although the above techniques allow for end-to-end optimization and have strong representation learning capabilities, automatically fusing multi-dimensional signals from data, most models still simplify the complex learner–resource–knowledge relationships into pairwise interactions, failing to explicitly model the prevalent high-order group behaviors and intricate knowledge topological structures. This limitation prompts the development of hypergraph-based recommender systems capable of effectively capturing complex high-order relations.

\subsection{Hypergraph-based recommendation techniques}
The hypergraph, a generalized graph structure that connects multiple nodes, can directly model high-order relationships between users and items \cite{sun2021heterogeneous}. Research indicates that preserving intact high-order relations, rather than breaking them down into pairwise relations, can improve downstream task performance \cite{lu2023schema}. This approach improves recommender system performance at the data representation level, rather than simply tuning the algorithms. As a result, researchers have created a set of hypergraph representation learning frameworks for encoding and propagating high-order associative information. Feng et al. developed a hypergraph neural network framework that uses hypergraph convolution to learn correlations in high-order data, proving its superiority in representing multi-modal data \cite{feng2019hypergraph}. Li et al. proposed a hypergraph representation learning model combining diffusion enhancement and hypergraph neural networks to improve anomaly detection capabilities \cite{li2025prototype}. Moreover, given the diversity of node and relation types in hypergraphs, heterogeneous hypergraph representation learning has rapidly gained attention. Related works include \cite{wang2024heterogeneous,bing2024efficient,zhang2025learning}, among others, which offer a more expressive framework to capture the rich, typed higher-order interactions of behavioral data.

Basically, hypergraph-based recommender systems are classified into two types based on their modeling complexity, i.e., Single-View Hypergraph-based Recommendation (SVHR) and Multi-View Hypergraph-based Recommendation (MVHR). SVHR typically models high-order relationships other than simple pairwise connections with a single dimension of relations. In social recommendation, for example, user group hyperedges can capture collaborative effects and interest propagation. A single hyperedge connects a group of users with strong social ties, thereby modeling complex social influences that transcend binary connections \cite{khan2025heterogeneous}. Chen et al. (2022) demonstrated that modeling user-group relationships can effectively address data sparsity issues by reformulating recommendation data into bipartite graphs and employing graph convolutional networks with attention mechanisms \cite{chen2022integrating}. Among aforementioned studies, the primary goal is to better understand and model the complex correlations found within a given data view. Many early SVHR models mainly focused on architectural innovations, such as denoising and data augmentation for hypergraphs \cite{wang2025hypergraph}, as well as enhancing the expressive power of hypergraph neural networks by increasing network depth or introducing more complex aggregation mechanisms \cite{li2025heterogeneous}. Recent studies during 2025 and 2026 years have significantly advanced SVHR by creating more sophisticated dynamic modeling mechanisms to capture the evolution of user preferences. Such dynamic hypergraph techniques can create hypergraph structures that change with user sessions or over time, capturing both dynamic social associations between users and high-order associations between items from multi-behavior interactions \cite{choi2025hypergraph}. For example, scholars have mapped user interaction behaviors to a latent intent space, considering both long-term repeated intent patterns and temporal decay of intents, thereby improving recommendation accuracy and diversity \cite{peintner2025hypergraph}. Further research suggests that focusing solely on long-term user preferences is insufficient and short-term interest drift should also be considered \cite{khan2025dynamic}. However, the significant limitation of SVHR is that all heterogeneous data is compressed into a single view for representation. When the intentions underlying user behaviors are multifaceted, a single view's modeling capacity becomes a bottleneck, resulting in insufficient and less robust learned representations.

In contrast, MVHR incorporates information from multiple heterogeneous sources, such as user-item interactions, item-item similarities, and cross-domain information constructed via meta-path hypergraphs, to create more holistic and robust user and item representations \cite{lyu2025multi}. Compared to SVHR, the primary advantage of MVHR is their hierarchical representation and deep fusion of complementary information from various perspectives. For example, in next point-of-interest (POI) recommendations, a user's decision is influenced by both their historical movement trajectory, i.e., temporal view and the spatial proximity between locations, i.e., spatial view \cite{lai2023multi}. Current MVHR research focuses on cross-view fusion \cite{zeng2025enhanced}, hyperedge prediction \cite{tian2026consistency}, and model generalizability optimization \cite{lin2025unified}. Some studies enforce consistency between embeddings of the same item across different views in the latent space to minimize distributional divergence. This enables cross-view contrastive learning for semantic alignment and adaptive fusion \cite{ma2024cross}. Other works incorporate self-supervised learning models to not only use explicit interaction signals but also mine deep, consistent feature representations from multi-view data through constructed contrastive tasks, thereby enhancing the model's generalization capability \cite{han2024dual}. To predict emerging high-order associations between nodes in dynamic networks, techniques such as dynamic temporal hypergraph representation have been proposed \cite{xu2025fast}. They generate corresponding hyperedge embeddings for candidate node sets and compute the probability of the hyperedge appearing in the future which efficiently and accurately predicts the potential for multiple nodes to co-participate in events within complex networks. Regarding model generalizability optimization, a common approach involves joint training with multiple optimization objectives, such as minimizing a main recommendation loss and a contrastive loss, to constrain the model to learn more generalizable and robust representations within a multi-task learning framework \cite{lin2025unified}.

Although MVHR integrates multi-source information and becomes a mainstream paradigm in the cutting-edge recommender systems, its applications are limited to education-decontextualized domains such as next POI recommendation \cite{lai2023multi}, session-based recommendation \cite{zeng2025enhanced}, Web API recommendation \cite{shen2024multi}, and news recommendation \cite{liu2024dual}. To date, mature applications of MVHR in PERS are conspicuously lacking, and the key challenge of how to effectively leverage multi-view information of hypergraphs in educational scenarios is still underexplored. Although some research has used hypergraph convolutional networks to capture many-to-many relationships between courses and collaborative learning patterns among learners \cite{su2025hypergraph}. It is essentially single-view model that fails to fully exploit and integrate multi-view information such as leaner-resource interactions, resource semantic associations, and learner knowledge states. Furthermore, PERS necessitates pedagogical logic alignment and modeling of dynamic behaviors. However, existing MVHR methods are difficult to apply directly to PERS because they lack explicit education-specific modeling mechanisms for the evolution of learner behaviors and semantic dependencies between learning resources. This makes hypergraph recommendation based on dynamic behavioral profiling a key means to crack the bottleneck of achieving fine-grained, pedagogy-aware personalization in adaptive learning environments.

\section{Preliminary}
\label{pl}
In PERS, learning interactions primarily occur between learners and learning resources. However, following the standard convention in recommender systems literature, we use the general terms \textit{user} and \textit{item} throughout this paper where \textit{user} corresponds to the learner and \textit{item} corresponds to the learning resource accessed by the learner.

\begin{definition}[User-Item Interaction]
\label{def:user-item-interaction}
Let \(\mathcal{U}\) and \(\mathcal{I}\) denote the sets of users and items, respectively. The interactions between users and items are defined as a binary relation represented by a set \(E \subseteq \mathcal{U} \times \mathcal{I}\), where each pair \((u, i) \in E\) indicates that user \(u\) has interacted with item \(i\). 
\end{definition}

A user may interact with multiple items, and an item may be interacted with by multiple users. We assume that every user has at least one interaction, i.e., for all \(u \in \mathcal{U}\), there exists an item \(i \in \mathcal{I}\) such that \((u, i) \in E\). Similarly, every item has at least one interaction, i.e., for all \(i \in \mathcal{I}\), there exists a user \(u \in \mathcal{U}\) such that \((u, i) \in E\).

\begin{definition}[Item-Category Mapping]
\label{def:item-category-mapping}
Let \(\phi: \mathcal{I} \to \mathcal{C}\) be the mapping function that assigns each item to its corresponding category. For an arbitrary item \(i \in \mathcal{I}\), \(\phi(i) \in \mathcal{C}\) denotes the unique category to which item \(i\) belongs. Conversely, a category may contain multiple items. For a given category \(c \in \mathcal{C}\), the set of items that belong to \(c\) is defined as \(\mathcal{I}_c = \{ i \in \mathcal{I} \mid \phi(i) = c \}\). We assume that every category contains at least one item, i.e., for all \(c \in \mathcal{C}\), \(|\mathcal{I}_c| \ge 1\).
\end{definition}

\begin{definition}[Interaction Bipartite Graph]
\label{def:interaction-bipartite-graph}
An interaction bipartite graph, denoted as \(\mathcal{G}_B = (\mathcal{V}_B, \mathcal{E}_B)\), is an undirected graph that models pairwise relationships between two disjoint sets of entities. The vertex set \(\mathcal{V}_B = \mathcal{U} \cup \mathcal{I}\) is partitioned into a user set \(\mathcal{U} = \{u_1, u_2, ..., u_{|\mathcal{U}|}\}\) and an item set \(\mathcal{I} = \{i_1, i_2, ..., i_{|\mathcal{I}|}\}\), with \(\mathcal{U} \cap \mathcal{I} = \emptyset\) and \(|\mathcal{V}_B| = |\mathcal{U}| + |\mathcal{I}| = n\). The edge set \(\mathcal{E}_B \subseteq \mathcal{U} \times \mathcal{I}\) consists of all observed interactions, where each edge \(e = (u, i) \in \mathcal{E}_B\) connects a user \(u \in \mathcal{U}\) to an item \(i \in \mathcal{I}\).
\end{definition}

Each node \(v \in \mathcal{V}_B\) is associated with a feature vector, forming the node feature matrix \(\mathbf{X} \in \mathbb{R}^{n \times d}\), where \(d\) is the dimensionality of the feature space. The topological structure of the bipartite graph can be described by an adjacency matrix 
$
\mathbf{A} = \begin{bmatrix}
\mathbf{0}_{|\mathcal{U}| \times |\mathcal{U}|} & \mathbf{B} \\
\mathbf{B}^{\top} & \mathbf{0}_{|\mathcal{I}| \times |\mathcal{I}|}
\end{bmatrix}
$, where the interaction matrix \(\mathbf{B} \in \{0, 1\}^{|\mathcal{U}| \times |\mathcal{I}|}\) is defined element-wise for \(u \in \mathcal{U}\) and \(i \in \mathcal{I}\) as:
\begin{equation}
    \mathbf{B}_{u,i} = 
    \begin{cases}
        1, & \text{if } (u, i) \in \mathcal{E}_B, \\
        0, & \text{otherwise}.
    \end{cases}
\end{equation}

\begin{definition}[Hypergraph]
\label{def:hypergraph}
A hypergraph is defined as a triplet \(\mathcal{G}_H = (\mathcal{V}_H, \mathcal{E}_H, \mathbf{W}_H)\), where \(\mathcal{V}_H = \{v_1, v_2, \ldots, v_{|\mathcal{V}_H|}\}\) is the vertex set and \(\mathcal{E}_H = \{e_1, e_2, \ldots, e_{|\mathcal{E}_H|}\}\) is the hyperedge set. A hyperedge \(e \in \mathcal{E}_H\) can connect more than two vertices, i.e., it is a non-empty subset of \(\mathcal{V}_H\): \(e \subseteq \mathcal{V}_H, e \neq \emptyset\). Each hyperedge \(e\) is assigned a non-negative weight, forming the diagonal hyperedge weight matrix \(\mathbf{W}_H \in \mathbb{R}^{|\mathcal{E}_H| \times |\mathcal{E}_H|}\), where \(\mathbf{W}_{H,ii} = w(e_i) \geq 0\).
\end{definition}

The structure of the hypergraph can be represented by an incidence matrix \(\mathbf{H} \in \{0,1\}^{|\mathcal{V}_H| \times |\mathcal{E}_H|}\), whose entry
$\mathbf{H}_{v,e} = 1\text{ if } v \in e$ and 0, \text{otherwise} indicates the membership of vertex \(v\) in hyperedge \(e\). The degree of a vertex \(v\), denoted as \(d(v) = \sum_{e \in \mathcal{E}_H} w(e) \cdot \mathbf{H}_{v,e}\), is defined as the sum of the weights of all hyperedges incident to it. The degree of a hyperedge \(e\), denoted as \(d(e) = \sum_{v \in \mathcal{V}_H} \mathbf{H}_{v,e}\), is defined as the number of vertices it connects.

\begin{definition}[Heterogeneous Hypergraph]
\label{def:heterogeneous-hypergraph}
A heterogeneous hypergraph is formally defined as a quintuple
$
\mathcal{G}_{HH} = (\mathcal{V}_{HH}, \mathcal{E}_{HH}, \mathcal{T}_v, \mathcal{T}_e, \mathbf{W}_{HH}),
$
where \(\mathcal{T}_v\) and \(\mathcal{T}_e\) denote the sets of predefined vertex types and hyperedge types, respectively. Each vertex \(v \in \mathcal{V}_{HH}\) is associated with a mapping function \(\phi_v: \mathcal{V}_{HH} \rightarrow \mathcal{T}_v\), and each hyperedge \(e \in \mathcal{E}_{HH}\) is associated with a mapping function \(\psi_e: \mathcal{E}_{HH} \rightarrow \mathcal{T}_e\). The hyperedge weight matrix is denoted as $\mathbf{W}_{HH} \in \mathbb{R}^{|\mathcal{E}_{HH}| \times |\mathcal{E}_{HH}|}$.
\end{definition}

The hypergraph is heterogeneous if \(|\mathcal{T}_v| + |\mathcal{T}_e| > 2\), indicating the presence of more than two types of objects. In the current research context, a heterogeneous hypergraph can be constructed with vertices of learners, learning resources, and categories that the learning resources belong. Hyperedges can then model various high-order relations, such as a student interacting with a set of learning resources, a set of learning resources belonging to the same category, or a group of learners exhibiting similar behavioral patterns. 

\section{Proposed method}
\label{pm}
\subsection{Overview}
\begin{figure*}[htb] % h=当前位置, t=页顶, b=页底, p=单独页
    \centering % 居中显示
    \includegraphics[width=0.8\textwidth]{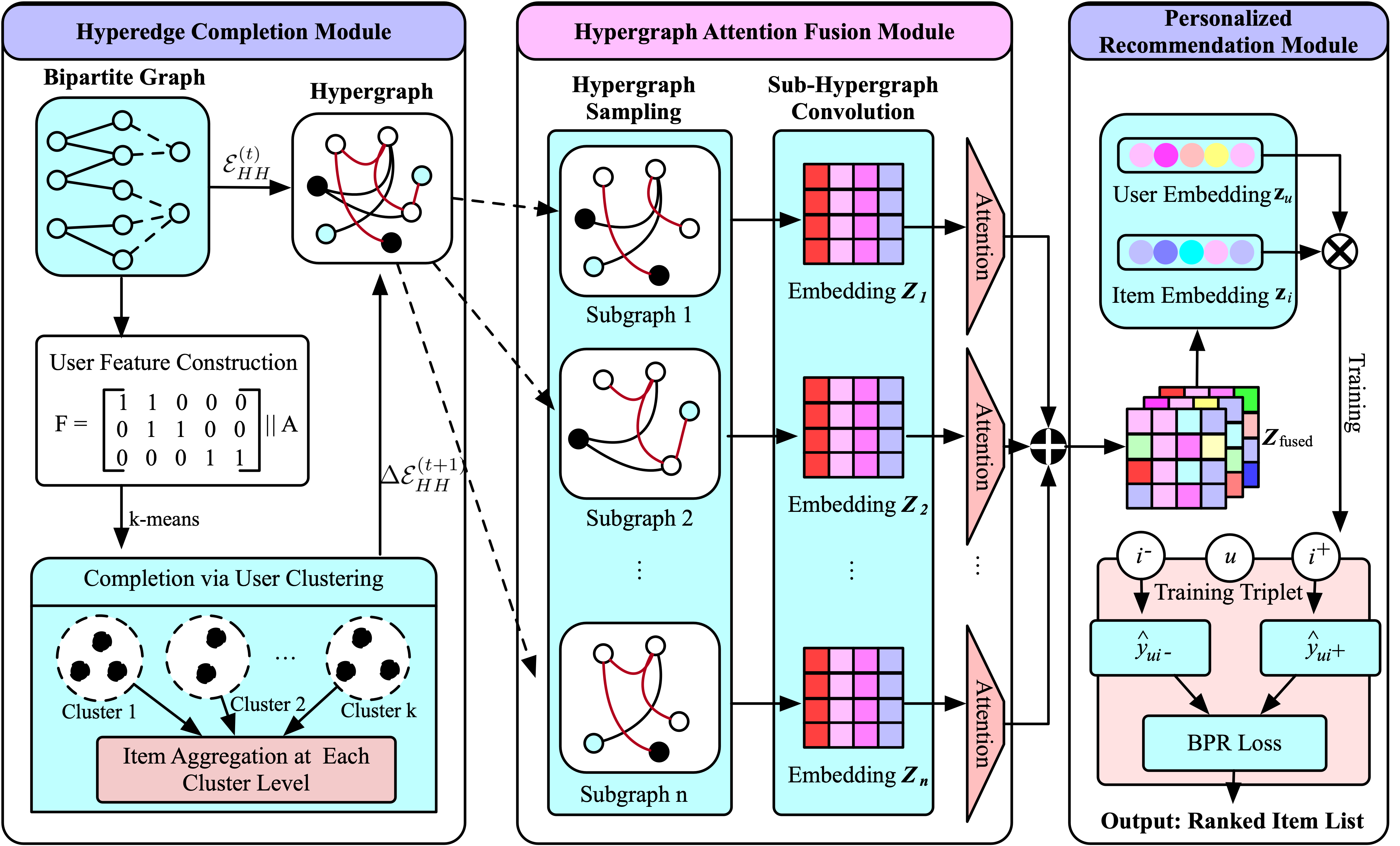} % 图片文件名(无需扩展名)
    \caption{Overall framework} % 图片标题
    \label{fig:figure1} % 唯一标识符(建议以fig:开头)
\end{figure*}

Building upon the established definitions, this research aims to develop an integrated pipeline that transforms observed user-item interactions and item-category mappings into a semantically-aware heterogeneous hypergraph, and ultimately learn predictive representations to deliver precise and educationally meaningful resource recommendations. The core input consists of a set of observed User-Item Interactions defined in Definition~\ref{def:user-item-interaction} and an Item-Category Mapping defined in Definition~\ref{def:item-category-mapping}. From this input, we first construct a hypergraph that seamlessly integrates users, items, and categories to capture a user’s holistic engagement pattern within a semantic context. This hypergraph is derived solely from explicit interactions. However, the potential high-order relationships, such as latent collaborative similarities among users with analogous behaviors, remain uncaptured. Therefore, we use a hypergraph completion step that infers a set of novel, high-quality candidate hyperedges to enrich the graph’s connectivity, thereby providing a more robust foundation for representation learning. The final output of this work is a ranked personalized recommendation list for each user. To achieve this, we learn dense, low-dimensional embedding vectors for all users and items, which encapsulate both their dynamic behavioral patterns and the multi-view relational structures encoded in the enriched hypergraph. The overall framework is illustrated in Figure \ref{fig:figure1}, which includes hyperedge construction  module detailed in Subsection~\ref{construt} and \ref{complete}, hypergraph information fusion module detailed in Subsection~\ref{fusion}, and personalized recommendation module detailed in Subsection~\ref{recommend}.

\subsection{Hyperedge construction}
\label{construt}
Given an interaction bipartite graph $\mathcal{G}_B$ (Definition~\ref{def:interaction-bipartite-graph}) and a category mapping function $\phi$ (Definition~\ref{def:item-category-mapping}), we construct a heterogeneous hypergraph $\mathcal{G}_{HH} = (\mathcal{V}_{HH}, \mathcal{E}_{HH}, \mathcal{T}_v, \mathcal{T}_e)$ by forming hyperedges between nodes of different types. Specifically, for each user $u \in \mathcal{U}$ and each category $c \in \mathcal{C}$, we define the set of items that $u$ has interacted with and that belong to $c$ as $\mathcal{I}_{u,c} = \{ i \in \mathcal{I} \mid (u, i) \in \mathcal{E}_B \ \land\ \phi(i) = c \}$. If $\mathcal{I}_{u,c} \neq \emptyset$, we construct a hyperedge $e$ containing a set of nodes $\{ u \} \cup \mathcal{I}_{u,c} \cup \{ c \}$. The set of all such hyperedges is $\mathcal{E}_{HH} = \left\{ \{ u \} \cup \mathcal{I}_{u,c} \cup \{ c \} \ \big|\ u \in \mathcal{U},\ c \in \mathcal{C},\ \mathcal{I}_{u,c} \neq \emptyset \right\}$. Formally, we define a composite hyperedge type as $t = (\mathcal{U}, \mathcal{I}, \mathcal{C}) \in \mathcal{T}_e$. A hyperedge $e$ of type $t$, denoted by $\psi_e = t$, encapsulagtes a high-order relation by connecting a specific user vertex $u \in \mathcal{U}$, a set of item vertices $\mathcal{I}_{u,c} \subseteq \mathcal{I}$, and a category vertex $c \in \mathcal{C}$ that semantically encompasses $\mathcal{I}_{u,c}$.

\begin{property}
\label{hyperedge-existence}
For any user $u \in \mathcal{U}$ and category $c \in \mathcal{C}$, a hyperedge $e$ exists in $\mathcal{E}_{HH}$ if and only if there exists at least one interaction in $\mathcal{G}_B$ between $u$ and some item $i$ where $\phi(i) = c$.
\end{property}

The proof of this property is obvious. (Sufficiency) If a hyperedge $e\in \mathcal{E}_{HH}$ containing nodes $\{u\} \cup \mathcal{I}_{u,c} \cup \{c\}$, then by construction, $\mathcal{I}_{u,c} \neq \emptyset$. By the definition of $\mathcal{I}_{u,c}$, there exists an item $i \in \mathcal{I}_{u,c}$ such that $(u, i) \in \mathcal{E}_B$ and $\phi(i)=c$. This directly proves the existence of such an interaction. (Necessity) If there exists an item $i$ such that $(u,i) \in \mathcal{E}_B$ and $\phi(i)=c$, then $i \in \mathcal{I}_{u,c}$, implying $\mathcal{I}_{u,c} \neq \emptyset$. By the hyperedge generation rule, a hyperedge $e$ for the pair $(u,c)$ is constructed and added to $\mathcal{E}_{HH}$.

Thus, for each user $u \in \mathcal{U}$ and each category $c \in \mathcal{C}$, the interaction set is constructed by $\mathcal{I}_{u,c} = \{ i \in \mathcal{I} \mid (u, i) \in \mathcal{E}_B \land \phi(i) = c \}$. A hyperedge $e$ is generated if and only if $\mathcal{I}_{u,c} \neq \emptyset$. The resulting hyperedge is a triple $(u, \mathcal{I}_{u,c}, c) \in \mathcal{E}_{HH}$. In this semantics, each hyperedge $e$ corresponds to a node set $\{u\} \cup \mathcal{I}_{u,c} \cup \{c\} \subseteq \mathcal{V}_{HH}$, where $\mathcal{V}_{HH} = \mathcal{U} \cup \mathcal{I} \cup \mathcal{C}$ is the vertex set of the heterogeneous hypergraph.

\begin{property}
\label{thm:hyperedge}
Let $\mathcal{G}_{HH} = (\mathcal{V}_{HH}, \mathcal{E}_{HH}, \mathcal{T}_v, \mathcal{T}_e)$ be a heterogeneous hypergraph constructed according to the hyperedge generation process. The cardinality of the hyperedge set $\mathcal{E}_{HH}$ satisfies $|\mathcal{E}_{HH}| \leq |\mathcal{U}| \times |\mathcal{C}|$. This bound is achievable when every user interacts with at least one item from each category.
\end{property}

The generation of hyperedges establishes an injective mapping from $\mathcal{E}_{HH}$ to the Cartesian product $\mathcal{U} \times \mathcal{C}$. Each hyperedge $e = (u, \mathcal{I}_{u,c}, c)$ is uniquely associated with the pair $(u, c)$. Formally, we define the mapping $\Psi: \mathcal{E}_{HH} \to \mathcal{U} \times \mathcal{C}$ by $\Psi(e) = (u, c)$ where each hyperedge corresponds to exactly one user-category pair. Since $\Psi$ is injective, we have $|\mathcal{E}_{HH}| \leq |\mathcal{U} \times \mathcal{C}| = |\mathcal{U}| \times |\mathcal{C}|$.

For each user $u \in \mathcal{U}$, the union of all item sets across hyperedges containing $u$ equals the set of items that $u$ interacted with in the original bipartite graph. Mathematically, we have $\bigcup_{c \in \mathcal{C}} \mathcal{I}_{u,c} = \{ i \in \mathcal{I} \mid (u, i) \in \mathcal{E}_B \}$. 

For example, consider a recommender system with $|\mathcal{U}| = 3$ users, $|\mathcal{I}| = 5$ items, and $|\mathcal{C}| = 2$ categories. The interaction bipartite graph $\mathcal{E}_B = \{(u_1, i_1), (u_1, i_2), (u_2, i_2), (u_2, i_3), (u_3, i_4), (u_3, i_5)\}$. The category mapping is $\phi(i_1) = \phi(i_2) = c_1$ and $\phi(i_3) = \phi(i_4) = \phi(i_5) = c_2$. Following the hyperedge generation process, we generate $\mathcal{I}_{u_1, c_1} = \{i_1, i_2\} \Rightarrow e_1 = (u_1, \{i_1, i_2\}, c_1)$, $\mathcal{I}_{u_2, c_1} = \{i_2\} \Rightarrow e_2 = (u_2, \{i_2\}, c_1)$, $\mathcal{I}_{u_2, c_2} = \{i_3\} \Rightarrow e_3 = (u_2, \{i_3\}, c_2)$, and $\mathcal{I}_{u_3, c_2} = \{i_4, i_5\} \Rightarrow e_4 = (u_3, \{i_4, i_5\}, c_2)$. Then, $\mathcal{E}_{HH} = \{e_1, e_2, e_3, e_4\}$ with $|\mathcal{E}_{HH}| = 4$, satisfying $|\mathcal{E}_{HH}| \leq |\mathcal{U}| \times |\mathcal{C}| = 6$.

By doing this, hyperedges can capture higher-order relationships among users, items, and categories, which explicitly encode semantic similarities to improve recommendation relevance. The pairwise interactions are aggregated into hyperedges to form a more compact graph topology, thereby reducing sparsity and mitigating computational overhead during model training.

\subsection{Hyperedge completion via dynamic behavioral profiling}
\label{complete}
The original hypergraph can establish high-order dependencies among multiple entities. However, it only includes observable interactions, resulting in insufficient representation of latent interactions. This  incomplete hypergraph leads to underutilization of information and restricts the ability to capture dynamic user hehaviors. To address this issue, hyperedge completion aims to extend the original hyperedge set to enhance the expressive power of the hypergraph. We propose a profiling-based hyperedge completion method. The core intuition is that users with similar feature representations and interaction patterns are likely to exhibit comparable preferences toward items. Through dynamic behavioral profiling, we update user profiles by analyzing real-time interactions and iteratively refining feature vectors. In practical implementation, we employ a clustering-based approach to group users with homogeneous behavioral patterns into hyperedges. By clustering such users, we can derive plausible hyperedges for individuals based on their group’s aggregated behavior, thereby enriching the hypergraph structure.

Let the original hyperedge set be $\mathcal{E}_{HH}^{(0)} = \{ e_1, e_2, \dots, e_m \}$, where each hyperedge $e = (u, \mathcal{I}_e, c)$ involves a user $u \in \mathcal{U}$, a set of items $\mathcal{I}_e \subseteq \mathcal{I}$, and a category label $c \in \mathcal{C}$. From $\mathcal{E}_{HH}^{(0)}$, we first extract the set of all unique users $\mathcal{U} = \{u_1, \dots, u_{|\mathcal{U}|}\}$, items $\mathcal{I} = \{i_1, \dots, i_{|\mathcal{I}|}\}$, and categories $\mathcal{C} = \{c_1, \dots, c_{|\mathcal{C}|}\}$. We then construct a user-category-item mapping $\Phi: \mathcal{U} \times \mathcal{C} \to 2^{\mathcal{I}}$ such that $\Phi(u, c)$ is the set of items in category $c$ that user $u$ has interacted with in $\mathcal{E}_{HH}^{(0)}$. Then, the hypergraph completion undergoes the following 4 steps. 

\begin{algorithm}
\caption{Hyperedge Completion}
\label{alg:hyperedge_extension}
\begin{algorithmic}[1]
\Require $\mathcal{E}_{HH}^{(0)}$, $\rho$, $K$
\Ensure $\mathcal{E}_{HH}^{(\text{com})}$

\State Build feature matrix $\mathbf{F}$ (Step 1)

\State $\{\mathcal{G}_1, \dots, \mathcal{G}_K\} \gets \text{clustering}(\mathbf{F}, K)$ (Step 2)
\State $g \gets$ cluster assignment function

\For{$k = 1$ to $K$, $c \in \mathcal{C}$}  (Step 3)
    \State $\mathcal{I}_{k,c} \gets \bigcup_{u \in \mathcal{G}_k} \Phi(u, c)$
\EndFor

\State $\mathcal{U}_{\text{sample}} \gets$ sample $\max(1, \lfloor \rho |\mathcal{U}| \rfloor)$ users from $\mathcal{U}$
\State $\Delta\mathcal{E} \gets \emptyset$
\For{$u \in \mathcal{U}_{\text{sample}}$}  (Step 4)
    \State $k \gets g(u)$
    \For{$c \in \mathcal{C}$ where $\mathcal{I}_{k,c} \neq \emptyset$}
        \State $\Delta \mathcal{E} \gets \Delta\mathcal{E} \cup \{(u, \mathcal{I}_{k,c}, c)\}$
    \EndFor
\EndFor
\State \Return $\mathcal{E}_{HH}^{(0)} \cup \Delta\mathcal{E}$
\end{algorithmic}
\end{algorithm}

\noindent \textbf{Step 1: Feature Construction.}
We construct a composite feature vector for each user by combining an interaction matrix and an auxiliary matrix. An interaction matrix $\mathbf{M} \in \{0,1\}^{|\mathcal{U}| \times |\mathcal{I}|}$ where $\mathbf{M}[p, q] = 1$ if user $u_p$ has interacted with item $i_q$ in any hyperedge within $\mathcal{E}_{HH}^{(0)}$, and $0$ otherwise. The indices $p$ and $q$ correspond to fixed orderings of $\mathcal{U}$ and $\mathcal{I}$. An auxiliary matrix $\mathbf{A} \in \mathbb{R}^{|\mathcal{U}| \times d_a}$ containing $d_a$-dimensional pre-existing features for each user (e.g., demographics, historical profiles). The final feature matrix $\mathbf{F} \in \mathbb{R}^{|\mathcal{U}| \times d_f}$ for all users can be represented as $\mathbf{F} = [\mathbf{A} \; \Vert \; \mathbf{M}]$, where $[\cdot \Vert \cdot]$ denotes horizontal concatenation and $d_f =  d_a + |\mathcal{I}|$. The $p$-th row of $\mathbf{F}$, denoted $\mathbf{f}_p \in \mathbb{R}^{d_f}$, is the feature vector for user $u_p$.

\noindent \textbf{Step 2: User Clustering.}
We partition the user set $\mathcal{U}$ into $K$ disjoint clusters $\{\mathcal{G}_1, \ldots, \mathcal{G}_K\}$ by applying the k-means algorithm to the feature vectors $\{\mathbf{f}_p\}_{p=1}^{|\mathcal{U}|}$. The objective is to minimize the within-cluster sum of squares:
\begin{align}
\min_{\{ \boldsymbol{\mu}_k \}_{k=1}^K, \{ z_{pk} \}} & \sum_{p=1}^{|\mathcal{U}|} \sum_{k=1}^{K} z_{pk} \, \| \mathbf{f}_p - \boldsymbol{\mu}_k \|^2 \\
\text{s.t. } & \sum_{k=1}^{K} z_{pk} = 1, \quad z_{pk} \in \{0, 1\}, \quad \forall p, k
\end{align}
where $\boldsymbol{\mu}_k \in \mathbb{R}^{d_f}$ is the centroid of cluster $k$, and $z_{pk}=1$ if and only if user $u_p$ is assigned to cluster $\mathcal{G}_k$. Let $g: \mathcal{U} \to \{1,\dots,K\}$ be the resulting cluster assignment function, i.e., $g(u_p) = k$ such that $z_{pk}=1$.

\noindent \textbf{Step 3: Cluster-Level Item Aggregation.}
For each cluster $\mathcal{G}_k$ and each category $c \in \mathcal{C}$, we compute the union of all items that users within the cluster have interacted with under that category using
$\mathcal{I}_{k, c} = \bigcup_{u \in \mathcal{G}_k} \Phi(u, c).$ This yields a cluster-category specific item set $\mathcal{I}_{k,c} \subseteq \mathcal{I}$.

\noindent \textbf{Step 4: Probabilistic Hyperedge Generation.}
To efficiently augment the hyperedge set, we employ a stochastic generation strategy. A subset of users $\mathcal{U}_{\text{sample}}$ is uniformly sampled at random from $\mathcal{U}$ without replacement, with size $|\mathcal{U}_{\text{sample}}| = \max\left(1, \lfloor \rho |\mathcal{U}| \rfloor \right)$, where $\rho \in (0, 1]$ is a sampling rate. For each sampled user $u \in \mathcal{U}_{\text{sample}}$ belonging to cluster $k = g(u)$, and for each category $c$ where $\mathcal{I}_{k,c} \neq \emptyset$, a new hyperedge $e_{\text{new}} = \big( u, \; \mathcal{I}_{k,c}, \; c \big)$ is generated.
The extended hyperedge set becomes $\mathcal{E}_{HH}^{(\text{com})} = \mathcal{E}_{HH}^{(0)} \cup \{ e_{\text{new}} \;|\; u \in \mathcal{U}_{\text{sample}}, \, c \in \mathcal{C}, \, \mathcal{I}_{g(u),c} \neq \emptyset \}$, which contains the original hyperedges supplemented with the generated group-based hyperedges $\Delta\mathcal{E}$.

\begin{property}
\label{thm:info-gain}
Let $\mathcal{E}_{HH}^{(0)}$ be the original hyperedge set and $\mathcal{E}_{HH}^{(\mathrm{com})}$ be the augmented set. The increase in hyperedge count satisfies $
|\mathcal{E}_{HH}^{(\mathrm{com})}| - |\mathcal{E}_{HH}^{(0)}| \leq |\mathcal{U}_{\mathrm{sample}}| \cdot |\mathcal{C}|$. The bound is tight if and only if: (i) $\mathcal{I}_{g(u),c} \neq \emptyset$ for every $u \in \mathcal{U}_{\mathrm{sample}}$ and $c \in \mathcal{C}$, and (ii) the hyperedge $\{u\} \cup \mathcal{I}_{g(u),c} \cup \{c\}$ is novel for all such pairs.
\end{property}

By Step~4, a new hyperedge is generated for a pair $(u, c)$ if and only if $u \in \mathcal{U}_{\mathrm{sample}}$ and $\mathcal{I}_{g(u),c} \neq \emptyset$. For each fixed $u \in \mathcal{U}_{\mathrm{sample}}$, at most $|\mathcal{C}|$ hyperedges can be created with one per category. Summing over all sampled users yields the bound $|\mathcal{U}_{\mathrm{sample}}| \cdot |\mathcal{C}|$. Condition (i) ensures that the algorithm attempts to generate a hyperedge for every pair $(u, c)$ with $u \in \mathcal{U}_{\mathrm{sample}}$ and $c \in \mathcal{C}$. Condition (ii) ensures that all $|\mathcal{U}_{\mathrm{sample}}| \cdot |\mathcal{C}|$ generated hyperedges are distinct from $\mathcal{E}_{HH}^{(0)}$ and from each other. 

The pseudo code is provided in Algorithm~\ref{alg:hyperedge_extension}. Based on the previous example, we now demonstrate how this algorithm augments $\mathcal{E}_{HH}^{(0)}$. Assume  there are no auxiliary features ($\mathbf{A} = \emptyset$) for simplicity, the feature matrix $\mathbf{F} = \mathbf{M}$ is constructed from interaction patterns. Applying k-means clustering with $K=2$, we obtain two clusters: $\mathcal{G}_1 = \{u_1, u_2\}$ and $\mathcal{G}_2 = \{u_3\}$. The cluster-level item aggregation yields $\mathcal{I}_{1, c_1} = \Phi(u_1, c_1) \cup \Phi(u_2, c_1) = \{i_1, i_2\} \cup \{i_2\} = \{i_1, i_2\}$, $\mathcal{I}_{1, c_2} = \Phi(u_1, c_2) \cup \Phi(u_2, c_2) = \emptyset \cup \{i_3\} = \{i_3\}$, $\mathcal{I}_{2, c_1} = \Phi(u_3, c_1) = \emptyset$, and $\mathcal{I}_{2, c_2} = \Phi(u_3, c_2) = \{i_4, i_5\}$. Setting a sampling rate $\rho = 1.0$ (i.e., all users are selected for hyperedge generation), we extend the hyperedge set as follows: For $u_1 \in \mathcal{G}_1$, we add $(u_1, \mathcal{I}_{1,c_1}, c_1) = (u_1, \{i_1, i_2\}, c_1)$ and $(u_1, \mathcal{I}_{1,c_2}, c_2) = (u_1, \{i_3\}, c_2)$; For $u_2 \in \mathcal{G}_1$, we add $(u_2, \mathcal{I}_{1,c_1}, c_1) = (u_2, \{i_1, i_2\}, c_1)$ and $(u_2, \mathcal{I}_{1,c_2}, c_2) = (u_2, \{i_3\}, c_2)$; For $u_3 \in \mathcal{G}_2$, we add $(u_3, \mathcal{I}_{2,c_2}, c_2) = (u_3, \{i_4, i_5\}, c_2)$ (no hyperedge for $c_1$ since $\mathcal{I}_{2,c_1} = \emptyset$). After removing duplicates, the extended hyperedge set becomes $\mathcal{E}_{HH}^{(\text{com})} = \mathcal{E}_{HH}^{(0)} \cup \left\{ (u_1, \{i_3\}, c_2), \; (u_2, \{i_1, i_2\}, c_1) \right\}$.

\subsection{Multi-view attention fusion}
\label{fusion}
\subsubsection{Hypergraph random walk sampling}
To capture diverse local structural characteristics of hypergraphs, we devise a random walk-based sampling strategy on the hypergraph topology, generating 
$m$ representative sub-hypergraph views $\left\{\mathcal{G}_{HH}^{(i)} = (\mathcal{V}_{HH}^{(i)}, \mathcal{E}_{HH}^{(i)}) \right\}_{i=1}^{m}$. The process is shown in Algorithm~\ref{alg:sampling}. The random walk on the hypergraph is defined by hyperedge selection $P(e \mid v) = \frac{w(e,v)}{\sum_{e' \in \mathcal{E}_{HH}(v)} w(e',v)}$ and node selection $P(v' \mid e, v) = \frac{w'(v',e)}{\sum_{u \in e \setminus \{v\}} w'(u,e)}$, where $w(e,v) = |e|$ is the weight of selecting hyperedge $e$ from node $v$, $w'(v',e) = d(v')$ is the weight of selecting node $v'$ from hyperedge $e$ where $d(v')$ is the degree of node $v'$. $\mathcal{E}_{HH}(v) = \{e \in \mathcal{E}_{HH} : v \in e\}$ is the set of hyperedges incident to $v$.

The walker restarts to the initial node $v_0$ with probability $\alpha$. The transition probability is calculated by:
\begin{equation}
P_{\text{next}}(v' \mid v) = \alpha \cdot \mathbb{I}_{v' = v_0} + (1 - \alpha) \cdot \sum_{e \in \mathcal{E}_{HH}(v)} P(e \mid v) \cdot P(v' \mid e, v),
\end{equation}
where $\mathbb{I}_{v' = v_0}$ is the indicator function that equals 1 if $v' = v_0$ and 0 otherwise. After $L$ steps, the visited nodes and hyperedges form a sub-hypergraph $\mathcal{G}_{HH}^{(i)} = (\mathcal{V}_{HH}^{(i)}, \mathcal{E}_{HH}^{(i)})$. The algorithm filters out sub-hypergraphs that are too small if and only if $|\mathcal{V}_{HH}^{(i)}| \geq \tau$, where $\tau$ is a predefined minimum node threshold. 

\begin{property}
\label{thm:expectation}
The approximate expected number of unique nodes in a sampled sub-hypergraph is:
\begin{equation}
\label{eq}
\mathbb{E}[|\mathcal{V}_{HH}^{(i)}|] = \sum_{v \in \mathcal{V}_{HH}} \left[1 - (1 - \pi_v)^L\right],
\end{equation}
where $\pi_v$ is the stationary probability of visiting node $v$ in the random walk.
\end{property}

Let $X_0, X_1, \dots, X_L$ denote the sequence of nodes visited in the random walk with restart, where $X_0 = v_0$ is the initial node and $L$ is the number of steps. For each node $v \in \mathcal{V}_{HH}$, define the indicator variable $I_v = \mathbb{I}\{ \exists t \in \{0,1,\dots,L\} : X_t = v \},$ which equals 1 if $v$ is visited at least once during the walk and 0 otherwise. The number of unique nodes in the sampled sub-hypergraph is $|\mathcal{V}_{HH}^{(i)}| = \sum_{v \in \mathcal{V}_{HH}} I_v$.
By linearity of expectation, 
\begin{equation}
\label{equa}
\mathbb{E}[|\mathcal{V}_{HH}^{(i)}|] = \sum_{v \in \mathcal{V}_{HH}} \mathbb{P}(I_v = 1).
\end{equation}

Assume the events $\{X_t = v\}$ across different steps are approximately independent, the probability that node $v$ is not visited in any of the $L$ steps is $\mathbb{P}(I_v = 0) \approx \prod_{t=1}^{L} \bigl(1 - \mathbb{P}(X_t = v)\bigr) \approx (1 - \pi_v)^L$. Consequently, we have:
\begin{equation}
\label{equaaa}
\mathbb{P}(I_v = 1) = 1 - \mathbb{P}(I_v = 0) \approx 1 - (1 - \pi_v)^L.
\end{equation}

Substituting Equation (\ref{equaaa}) into Equation (\ref{equa}), the Equation (\ref{eq}) can be obtained.

\begin{algorithm}
\caption{Hypergraph Sampling}
\label{alg:sampling}
\begin{algorithmic}[1]
\Require $\mathcal{G}_{HH} $, $m$, $L$, $\alpha$
\Ensure Set of sub-hypergraphs $\mathcal{S}$
\State $\mathcal{S} \leftarrow \emptyset$
\For{$i = 1$ \textbf{to} $m$}
    \State Randomly select a starting node $v_0 \in \mathcal{V}_{HH}$
    \State $\mathcal{V}_{HH}^{(i)} \leftarrow \{v_0\}$, $\mathcal{E}_{HH}^{(i)} \leftarrow \emptyset$, $v \leftarrow v_0$
    \For{$t = 1$ \textbf{to} $L$}
        \If{$\text{rand}() < \alpha$}
            \State $v \leftarrow v_0$
        \Else
            \State Sample hyperedge $e$ from $\mathcal{E}_{HH}(v)$ with probability $P(e \mid v)$
	   \State $\mathcal{E}_{HH}^{(i)} \gets \mathcal{E}_{HH}^{(i)} \cup \{e\}$
           \State Sample node $v'$ from $e \setminus \{v\}$ with probability $P(v' \mid e, v)$
           \State $\mathcal{V}_{HH}^{(i)} \gets \mathcal{V}_{HH}^{(i)} \cup \{v'\}$
            \State $v \leftarrow v'$
        \EndIf
    \EndFor
    
    \If{$|\mathcal{V}_{HH}^{(i)}| \geq \tau$}
        \State $\mathcal{S} \leftarrow \mathcal{S} \cup \{\mathcal{G}_{HH}^{(i)}\}$ where $\mathcal{G}_{HH}^{(i)} = (\mathcal{V}_{HH}^{(i)}, \mathcal{E}_{HH}^{(i)})$
    \EndIf
\EndFor
\State \Return $\mathcal{S}$
\end{algorithmic}
\end{algorithm}

\subsubsection{Sub-hypergraph convolution}
Traditional graph convolutional networks are limited to modeling pairwise relationships between nodes, which can not capture the higher-order interactions among users, items, and categories. To address this limitation, we employ hypergraph convolutional operation which consists of four components. First,  we initiate the process with hyperedge feature aggregation, formulated as $A = H^T X^{(l)},$ where $H \in \{0,1\}^{|\mathcal{V}_{HH}| \times |\mathcal{E}_{HH}|}$ denotes the incidence matrix and $X^{(l)} \in \R^{|\mathcal{V}_{HH}| \times d_l}$ represents the node features at layer $l$. This operation aggregates the features of all nodes within each hyperedge, effectively capturing the collective characteristics of nodes participating in the same higher-order relationship. The aggregated hyperedge features then undergo nonlinear transformation using $B = \sigma\left(A W_{\text{edge}}^{(l)}\right)$, where $W_{\text{edge}}^{(l)} \in \R^{d_l \times d_{l+1}}$ is the learnable weight matrix for hyperedge transformation, and $\sigma$ denotes the activation function. This transformation allows the model to learn semantic representations at the hyperedge level, automatically distinguishing between different types of interactions. The transformed hyperedge features are subsequently propagated back to nodes through node feature aggregation with $N = H B$.

We combine the node's intrinsic features with the aggregated neighborhood information through a residual connection:
\begin{equation}
X^{(l+1)} = \text{ReLU}\left(X^{(l)} W_{\text{node}}^{(l)} + N\right),
\end{equation}
where $W_{\text{node}}^{(l)} \in \R^{d_l \times d_{l+1}}$ is the learnable weight matrix for node feature transformation. For an $L$-layer hypergraph convolutional network, the final node embeddings are computed as $Z_i = \text{GNN}\left(X^{(0)}, H; \Theta\right)$, where $\Theta = \left\{W_{\text{edge}}^{(l)}, W_{\text{node}}^{(l)}\right\}_{l=0}^{L-1}$ represents all learnable parameters.

\subsubsection{Attention fusion}
Given $n$ node embeddings $\{\bm{Z}_1, \bm{Z}_2, \ldots, \bm{Z}_n\}$, where $\bm{Z}_i \in \R^{|\mathcal{V}_{HH}| \times d}$ represents the node embeddings learned from the $i$-th sub hypergraph, we employ an attention-based fusion mechanism to integrate these multi-view representations. For each sub hypergraph embedding $\bm{Z}_i$, we first compute a global query vector with mean pooling that serves as a summary representation of the entire sub hypergraph:

\begin{equation}
\bm{q}_i = \frac{1}{|\mathcal{V}_{HH}|} \sum_{j=1}^{|\mathcal{V}_{HH}|} \bm{Z}_i^{(j)},
\label{eq:query_vector}
\end{equation}
where $\bm{Z}_i^{(j)} \in \R^d$ denotes the embedding vector of the $j$-th node in the $i$-th sub hypergraph. This global query vector $\bm{q}_i \in \R^d$ encapsulates the overall characteristics of the sub hypergraph.

Then, we compute attention scores using:
\begin{equation}
s_i = \bm{W}_a \cdot [\overline{\bm{q}}_i \| \bm{Z}_i] + b_a,
\end{equation}
where $\bm{W}_a \in \R^{1 \times 2d}$ is the learnable attention weight matrix, $b_a \in \R$ is the bias term, $[\cdot \| \cdot]$ denotes the concatenation operation along the feature dimension, and $\overline{\mathbf{q}}_i \in \mathbb{R}^{|\mathcal{V}_{HH}| \times d}$ is the matrix obtained by repeating the row vector $\mathbf{q}_i^\top$ across all $|\mathcal{V}_{HH}|$ nodes. To ensure proper weighting across different sub hypergraph views, we apply the softmax function to normalize the attention scores $\alpha_i = \frac{\exp(s_i)}{\sum_{j=1}^n \exp(s_j)}$ for each node across different sub hypergraphs. 

The fused embedding is obtained through a weighted combination of the sub hypergraph embeddings using the computed attention weights:
\begin{equation}
\bm{Z}_{\text{fused}} = \sum_{i=1}^n \alpha_i \odot \bm{Z}_i,
\end{equation}
where $\odot$ denotes element-wise multiplication implemented via broadcasting. The fused embeddings undergo a final linear transformation to enhance their representational capacity using $\bm{Z}_{\text{out}} = \bm{W}_{\text{linear}} \bm{Z}_{\text{fused}} + \bm{b}_{\text{linear}}$, where $\bm{W}_{\text{linear}} \in \R^{d \times d}$ and $\bm{b}_{\text{linear}} \in \R^d$ are learnable parameters. 

\subsection{Personalized recommendation}
\label{recommend}
The final prediction layer combines the learned node embeddings with bias terms to generate preference scores. For a given user-item pair $(u, i)$ where $u \in \mathcal{U}$ and $i \in \mathcal{I}$, the prediction is computed as $\hat{y}_{ui} = \sigma\left( \mathbf{z}_u^\top \mathbf{z}_i + b_u + b_i \right)$, where $\mathbf{z}_u, \mathbf{z}_i \in \mathbb{R}^d$ are the fused embeddings for user $u$ and item $i$ respectively, $b_u$ and $b_i$ are learnable bias terms, and $\sigma(\cdot)$ denotes the sigmoid function.

To optimize the model parameters, we employ the Bayesian Personalized Ranking (BPR) loss, which is particularly suitable for implicit feedback scenarios. The loss function is defined as:

\begin{equation}
\mathcal{L}_{\text{BPR}} = -\frac{1}{|\mathcal{D}|} \sum_{(u,i^+,i^-) \in \mathcal{D}} \ln \sigma(\hat{y}_{ui^+} - \hat{y}_{ui^-}) + \lambda \|\Theta\|^2,
\label{eq:bpr_loss}
\end{equation}
where $\mathcal{D} = \{(u,i^+,i^-)\}$ represents the set of training triplets, with $i^+$ denoting positive items (interacted) and $i^-$ denoting negative items (non-interacted) for user $u$. The regularization term $\lambda \|\Theta\|^2$ prevents overfitting, with $\lambda$ controlling the regularization strength.

\begin{property}
The time complexity of proposed model is $O(kL|\mathcal{V}_{HH}|\bar{d}d)$, where $k$ is the number of sub-hypergraphs, $L$ is the number of GCN layers, $|\mathcal{V}_{HH}|$ is the number of nodes, $\bar{d}$ is the average node degree, and $d$ is the embedding dimension.
\end{property}

The random walk process for each sub-hypergraph requires $O(L_w)$ operations, where $L_w$ is the walk length. For $k$ sub-hypergraphs, this yields $O(kL_w)$. Each convolution layer of hypergraph convolution module involves edge feature aggregation: $O(|\mathcal{E}_{HH}|d)$, edge feature transformation: $O(|\mathcal{E}_{HH}|d^2)$, node feature aggregation: $O(|\mathcal{V}_{HH}|\bar{d}d)$, and node feature transformation: $O(|\mathcal{V}_{HH}|d^2)$. Since $|\mathcal{E}_{HH}| \leq |\mathcal{V}_{HH}|\bar{d}$, the dominant term is $O(|\mathcal{V}_{HH}|\bar{d}d^2)$ per layer. For $L$ layers and $k$ sub-hypergraphs, the total cost is $O(kL|\mathcal{V}_{HH}|\bar{d}d^2)$. The attention fusion step requires $O(k|\mathcal{V}_{HH}|d)$ operations for query generation and $O(k|\mathcal{V}_{HH}|d^2)$ for attention scoring.

Considering that $d$ is typically small (e.g., 64-256) and $L_w \ll |\mathcal{V}_{HH}|$, the overall time complexity is dominated by the graph convolution operations, yielding $O(kL|\mathcal{V}_{HH}|\bar{d}d^2)$. Because $d$ is fixed and small in practice, we simplify the time complexity to $O(kL|\mathcal{V}_{HH}|\bar{d}d)$. Moreover, when $k$ and $d$ are treated as constants, the model has linear time complexity $O(|\mathcal{V}_{HH}|\bar{d})$ with respect to the graph size.

\section{Experimental procedure}
\label{ep}
\subsection{Setup}
All experiments were conducted on a uniform computing platform equipped with two NVIDIA A30 GPUs of 24 GB memory each. The software environment was standardized using Python 3.12.4 and PyTorch 2.4.0 compiling against CUDA 12.1. The datasets were randomly partitioned into training, validation, and test sets in the ratio of \SI{80}{\percent}, \SI{10}{\percent}, and \SI{10}{\percent}, respectively. All experiments were conducted over 10 iterations, with mean values reported. For Top-$K$ recommendations, $K$ was set to 5, 10, 15, and 20. In random walks, the restart probability was 0.1, the number of walk steps was 15, and the number of sub-hypergraphs was 5. The learning rate was 0.001, the embedding dimension was 64, the regularization coefficient was 0.00001, and the optimizer used was Adam.

\subsection{Datasets}
We employ five publicly available and one self-constructed datasets to evaluate our proposed framework. The datasets vary in scale, domain, and structural characteristics, providing a comprehensive testbed for our model. Key statistics are summarized in Table \ref{tab:dataset-statistics}.

\begin{itemize}
    \item \textbf{ASSISTments2009}~\cite{feng2009addressing}. 
    This dataset consists of student interaction logs from the ASSISTments blended learning platform during the 2009–2010 academic year. It includes three entity types, i.e., students, problems, and knowledge components, along with their corresponding interaction records, capturing the platform's tutoring-as-assessment design.

    \item \textbf{ASSISTments2017}~\cite{patikorn2020assistments}.
    Collected between 2004 and 2007 and released for the 2017 ASSISTments data mining competition, this dataset extends ASSISTments2009 with richer student attributes such as average knowledge state, engagement level, and confusion metrics. These additions support a more fine-grained analysis of learning behavior.

    \item \textbf{MOOCCubeX}~\cite{yu2021mooccubex}.
    This dataset is a large-scale, knowledge-centered repository for adaptive learning in Massive Open Online Courses (MOOCs), sourced from the XuetangX platform. It records video-watching behaviors from 2020, encompassing 4,216 courses, 230,263 video learning resources, 358,265 exercises, 637,572 fine-grained concepts, and over 296 million raw behavioral records from 3,330,294 learners. This dataset is notable for its massive scale and fine-grained knowledge decomposition.

    \item \textbf{MOOPer}~\cite{liu2021mooper}.
    This dataset was derived from EduCoder, a large-scale online practice-oriented learning platform for information technology majors, covering data from 2018--2019. It includes entities such as learners, courses, practices, difficulty levels, and knowledge points, with diverse interactions among them. This dataset is characterized by its focus on hands-on programming and project-based learning.

    \item \textbf{CiteULike-t}~\cite{wang2013collaborative}.
    Built from CiteULike, a free reference management service, this dataset records user--paper interactions and paper--tag associations, forming a widely used benchmark for collaborative tagging and tag recommendation research.

    \item \textbf{Self-Build Dataset}.
    To further assess the generalizability of our approach, we constructed a new dataset from a provincial K--12 education platform in western China, with data collected up to March 2, 2023. Covering the 2020--2023 school years, it includes entities for learners, knowledge points, and subjects, along with relations that capture knowledge mastery and knowledge‑point--subject mappings, offering a realistic testbed in a formal school setting.
\end{itemize}

\begin{table*}[htb]
\centering
\small
\caption{Statistics of the datasets.}
\label{tab:dataset-statistics}
\begin{tabular}{lrrrrrr}
\toprule
Dataset & \makecell{\#Interactions} & \makecell{\#Vertices} & \makecell{\#Hyperedges} & \makecell{Avg.\\Hyperedge\\Degree} & \makecell{Max.\\Hyperedge\\Degree} & 
\makecell{Min.\\Hyperedge\\Degree} \\
\midrule
Assistment2009 & 335,394 & 29,008 & 36,187 & 11.78 & 500 & 2 \\
Assistment2017 & 390,331 & 4,399 & 89,642 & 16.51 & 883 & 2 \\
CiteULike-t & 123,896 & 25,586 & 64,735 & 38.04 & 2,490 & 2 \\
MOOCCubeX & 49,324 & 2,682 & 258,384 & 7.36 & 92 & 2 \\
MOOPer & 181,366 & 5,341 & 88,804 & 4.24 & 38 & 2 \\
SelfBuild & 936,640 & 5,399 & 13,165 & 80.65 & 591 & 2 \\
\bottomrule
\end{tabular}
\end{table*}

\subsection{Baseline methods}
To evaluate the effectiveness of our proposed method, we compare it with the following state-of-the-art baseline algorithms:

\begin{itemize}

    \item \textbf{Hypergraph Neural Networks (HGNN)}~\cite{feng2019hypergraph}. HGNN is a general framework designed to encode high-order correlations in data by leveraging hypergraph structures. It uses a hyperedge convolution mechanism to effectively capture complex data associations, making it particularly advantageous for multimodal data representation learning.
    
    \item \textbf{Heterogeneous Graph Propagation Network (HPN)}~\cite{ji2021heterogeneous}. HPN is a novel propagation network that addresses semantic confusion in heterogeneous graph representation learning. It enhances node-level aggregation through a semantic propagation mechanism that fuses local node semantics with adaptive weights. Furthermore, it employs a semantic fusion mechanism to learn the importance of different meta-paths and integrate them judiciously.

    \item \textbf{Hypergraph Wavelet Neural Network (HWNN)}~\cite{sun2021heterogeneous}. HWNN is a representation learning framework based on graph neural networks. It first projects the heterogeneous hypergraph into a series of snapshots, then performs local hypergraph convolution using wavelet bases derived from polynomial approximations, thereby avoiding the computationally expensive Laplacian decomposition required by traditional Fourier-based methods. This approach is effective for tasks involving both pairwise and non-pairwise relations.

    \item \textbf{Heterogeneous Hypergraph Variational Autoencoder (HeteHG)}~\cite{fan2021heterogeneous}.  HeteHG first maps a conventional heterogeneous information network into a semantically-rich heterogeneous hypergraph to capture high-order semantics and complex relations while preserving low-order pairwise topology. It then learns deep latent representations of nodes and hyperedges in an unsupervised manner via a Bayesian deep generative framework. An additional hyperedge attention module is incorporated to learn the importance of different node types within each hyperedge, enabling the modeling of multi-level relations in heterogeneous environments.

    \item \textbf{Self-Attention based Graph Neural Network (SAGNN)}~\cite{zhang2019hyper}. SAGNN generates static node embeddings via position-aware feedforward networks and employs multi-head attention mechanisms to dynamically aggregate neighborhood features within hyperedges. This design allows it to handle both homogeneous and heterogeneous hypergraphs with variable hyperedge sizes.

    \item \textbf{Hypergraph Joint Representation Learning (HJRL)}~\cite{yan2024hypergraph}. HJRL transforms both nodes and hyperedges into nodes within an expanded graph, establishing connections among node-node, hyperedge-hyperedge, and node-hyperedge pairs. Through a joint learning framework, it co-embeds nodes and hyperedges into a unified representation space to extract common knowledge. HJRL effectively preserves the rich relational information in hypergraphs, thereby enhancing both information aggregation and graph reconstruction capabilities.
\end{itemize}

\subsection{Evaluation metrics}
To comprehensively evaluate the effectiveness of the proposed model, we employ five widely-adopted metrics in recommender systems: Precision (P), Recall (R), Mean Reciprocal Rank (MRR), normalized Discounted Cumulative Gain (nDCG), and the F1-score. For ranking-based evaluation, metrics are computed at a cut-off denoted as @$K$, representing the length of the recommendation list.

\begin{itemize}
    \item P@$K$ measures the fraction of relevant items within the top-$K$ recommended list:
    \begin{equation}
    	\text{P@}K = \frac{1}{|U|} \sum_{u \in U} \frac{|R^K(u) \cap T(u)|}{|R^K(u)|},
    \end{equation}
    where $U$ is the set of users, $R^K(u)$ denotes the top-$K$ recommended items for user $u$, and $T(u)$ represents the set of items that user $u$ has interacted with in the test set.

    \item R@$K$ quantifies the fraction of a user's relevant items that are successfully retrieved in the top-$K$ list:
    \begin{equation}
    	\text{R@}K = \frac{1}{|U|} \sum_{u \in U} \frac{|R^K(u) \cap T(u)|}{K}.
    \end{equation}

    \item MRR@$K$ evaluates the ranking quality by averaging the reciprocal rank of the first relevant item in the recommendation list for each user:
    \begin{equation}
    	\text{MRR@}K = \frac{1}{|U|} \sum_{u \in U} \frac{1}{\text{rank}_u}.
    \end{equation}
    Here, $\text{rank}_u$ is the position of the first relevant item for user $u$ in the ranked list. If no relevant item is found, $\frac{1}{\text{rank}_u}$ is set to 0. MRR ranges from 0 to 1, with higher values indicating better performance.

    \item nDCG@$K$ assesses the ranking quality by considering both the relevance and the position of items. It is defined as:
    \begin{equation}
    	\text{nDCG@}K = \frac{1}{|U|} \sum_{u \in U} \frac{\sum_{i=1}^{K} \frac{\mathbb{I}(R_i^K(u) \in T(u))}{\log_2(i + 1)}}{\sum_{i=1}^{min(K,|T(u)|} \frac{1}{\log_2(i + 1)}},
    \end{equation}
    where $R_i^K(u)$ is the item at position $i$ in the top-$K$ recommendation list for user $u$, and $\mathbb{I}(\cdot)$ is an indicator function that equals 1 if the condition is true and 0 otherwise. The denominator represents the ideal DCG at K, ensuring the metric is normalized between 0 and 1.

    \item F1-score provides a single metric that balances Precision and Recall by computing their harmonic mean:
    \begin{equation}
    	\text{F1@}K = 2 \cdot \frac{\text{P@}K \times \text{R@}K}{\text{P@}K + \text{R@}K}.
    \end{equation}
\end{itemize}

\section{Experimental results}
\label{er}
\subsection{Overall performance}
The experimental results are summarized in Table \ref{tab:performance_comparison}, with the best performance on each metric highlighted in bold. Our proposed model achieves the best or second-best performance across all evaluated datasets. It outperforms all baseline models on most metrics for the Assistment2009, Assistment2017, MOOCCubeX, and MOOPer datasets. Although the HWNN and HJRL models show slightly better results on the CiteULike-t and SelfBuild datasets, our model's performance remains very close to these top scores, confirming its robust and state-of-the-art capability.

Notably, a clear trend is observed between the hyperedge completion ratio and model improvement. As shown in Table \ref{tab:performance_comparison}, for all datasets except Assistment2009, a higher completion ratio generally corresponds to better performance across metrics. This finding reinforces the value of the hypergraph completion mechanism in enriching structural connectivity and strengthening the model's representational power.

\begin{table*}[htb]
\centering
\small
\caption{Performance comparison}
\label{tab:performance_comparison}
\begin{tabular}{@{} l c *{7}{c} *{3}{c} @{}}
\toprule
\multirow{2}{*}{Datasets} & \multirow{2}*{Metrics} & \multicolumn{6}{c}{Models} & \multicolumn{3}{c}{Ours} \\
\cmidrule(lr){3-8}
\cmidrule(lr){9-11}
& & {HGNN} & {HPN} & {HWNN} & {HeteHG} & {SAGNN} & {HJRL} & {$\rho = 0\%$} & {$\rho = 1\%$} & {$\rho = 5\%$} \\
\midrule
\multirow{4}{*}{Assistment2009} 
& P@10    & 0.3589 & 0.3424 & 0.3371 & 0.3607 & 0.3349 & 0.3915 & \textbf{0.4541} & 0.4514 & 0.4505 \\
& R@10    & 0.5326 & 0.4929 & 0.4707 & 0.5427 & 0.4566 & 0.6335 & \textbf{0.7935} & 0.7566 & 0.6230 \\
& NDCG@10 & 0.6640 & 0.6176 & 0.6357 & 0.6565 & 0.5902 & 0.7279 & \textbf{0.8722} & 0.8501 & 0.7801 \\
& MRR@10  & 0.5045 & 0.4405 & 0.4806 & 0.4846 & 0.4348 & 0.5727 & \textbf{0.7963} & 0.7736 & 0.6796 \\
\midrule
\multirow{4}{*}{Assistment2017} 
& P@10    & 0.8004 & 0.8014 & 0.7942 & 0.8018 & 0.8008 & 0.8005 & 0.8273 & 0.8313 & \textbf{0.8375} \\
& R@10    & 0.3905 & 0.3909 & 0.3786 & 0.3918 & 0.3906 & 0.3908 & 0.4189 & \textbf{0.4143} & 0.3907 \\
& NDCG@10 & 0.9271 & 0.9276 & 0.9217 & 0.9279 & 0.9267 & 0.9259 & 0.9496 & \textbf{0.9512} & 0.9485 \\
& MRR@10  & 0.8734 & 0.8769 & 0.8703 & 0.8759 & 0.8728 & 0.8655 & \textbf{0.9183} & 0.9158 & 0.9091 \\
\bottomrule
\multirow{4}{*}{MOOCCubeX} 
& P@10    & 0.3241 & 0.2770 & 0.2748 & 0.3169 & 0.3073 & 0.3302 & 0.4225 & 0.4367 & \textbf{0.4814} \\
& R@10    & 0.5542 & 0.4896 & 0.4964 & 0.5158 & 0.5035 & 0.5679 & 0.8078 & 0.8326 & \textbf{0.8619} \\
& NDCG@10 & 0.6529 & 0.6506 & 0.6570 & 0.6163 & 0.6206 & 0.6572 & 0.8337 & 0.8468 & \textbf{0.8534} \\
& MRR@10  & 0.5002 & 0.4721 & 0.4700 & 0.4975 & 0.4779 & 0.5359 & 0.7317 & 0.7470 & \textbf{0.7588} \\
\midrule
\multirow{4}{*}{MOOPer} 
& P@10    & 0.1747 & 0.1682 & 0.1975 & 0.1740 & 0.1861 & 0.1761 & 0.3615 & 0.3701 & \textbf{0.3857} \\
& R@10    & 0.4022 & 0.4027 & 0.4854 & 0.3821 & 0.4425 & 0.3872 & \textbf{0.9224} & 0.9153 & 0.8823 \\
& NDCG@10 & 0.4976 & 0.5286 & 0.5840 & 0.4714 & 0.5364 & 0.4758 & 0.9048 & 0.9057 & \textbf{0.9071} \\
& MRR@10  & 0.3335 & 0.3364 & 0.3789 & 0.3291 & 0.3652 & 0.3368 & 0.8718 & 0.8751 & \textbf{0.8773} \\
\midrule
\multirow{4}{*}{CiteULike-t} 
& P@10    & 0.9472 & 0.9453 & \textbf{0.9482} & 0.9469 & 0.9480 & 0.9446 & 0.9448 & 0.9477 & 0.9442 \\
& R@10    & 0.2462 & 0.2443 & 0.2472 & 0.2454 & \textbf{0.2487} & 0.2446 & 0.2458 & \textbf{0.2487} & 0.2398 \\
& NDCG@10 & 0.9840 & 0.9846 & 0.9834 & 0.9848 & 0.9848 & \textbf{0.9858} & 0.9821 & 0.9833 & 0.9830 \\
& MRR@10  & 0.9724 & 0.9784 & 0.9727 & 0.9770 & 0.9742 & \textbf{0.9842} & 0.9699 & 0.9699 & 0.9720 \\
\midrule
\multirow{4}{*}{SelfBuild} 
& P@10    & 0.8757 & 0.8655 & 0.8957 & \textbf{0.9210} & 0.9167 & 0.8956 & 0.9143 & 0.9140 & 0.9152 \\
& R@10    & 0.2447 & 0.2372 & 0.2621 & \textbf{0.2893} & 0.2875 & 0.2633 & 0.2823 & 0.2806 & 0.2706 \\
& NDCG@10 & 0.9561 & 0.9475 & 0.9633 & \textbf{0.9678} & 0.9569 & 0.9648 & 0.9667 & 0.9666 & 0.9655 \\
& MRR@10  & 0.9561 & 0.9475 & \textbf{0.9633} & 0.9678 & 0.9569 & 0.9648 & 0.9373 & 0.9374 & 0.9373 \\
\bottomrule
\end{tabular}%
\end{table*}

\subsection{Influence of recommendation length $K$}
\begin{figure*}[htb] % h=当前位置, t=页顶, b=页底, p=单独页
    \centering % 居中显示
    \includegraphics[width=0.8\textwidth]{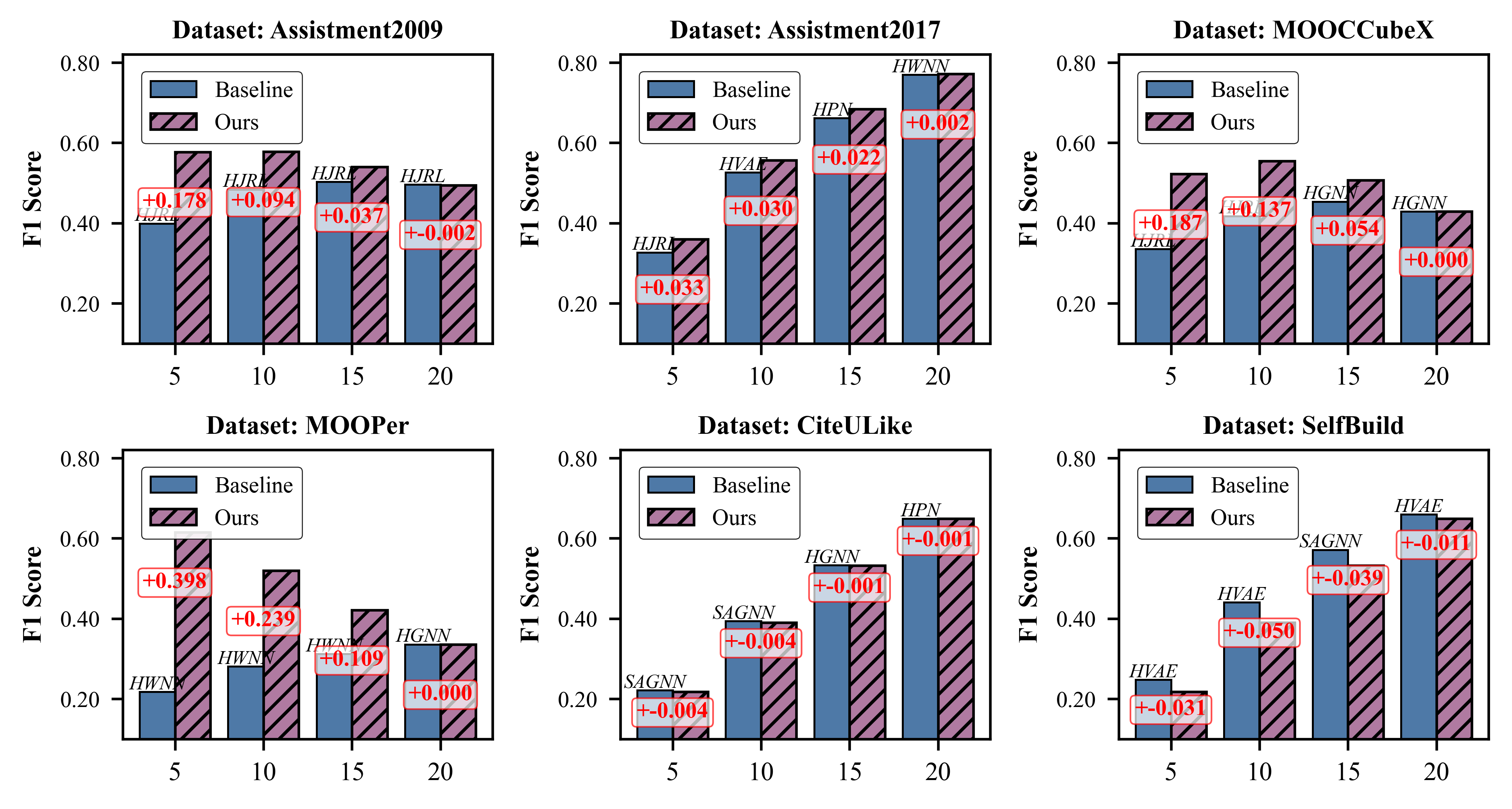} % 图片文件名(无需扩展名)
    \caption{F1 scores versus $K$ between the best baseline and ours} % 图片标题
    \label{length} % 唯一标识符(建议以fig:开头)
\end{figure*}

As shown in Figure \ref{length}, our model achieves a higher F1-score than the best-performing baseline across most datasets. This improvement is particularly significant at smaller recommendation list lengths ($K$ = 5, 10, 15). The most notable gain occurs at $K$ = 5 where our model achieves a 39.8\% relative increase. As $K$ increases to 20, the F1 scores of our model and the top baseline converge to nearly identical levels. This pattern indicates that our model holds particular advantage for low-rank recommendation scenarios, which is especially valuable for real-world applications that prioritize high precision in top-$K$ recommendations.

\subsection{Influence of hypergraph complexity}
\begin{figure*}[htb] % h=当前位置, t=页顶, b=页底, p=单独页
    \centering % 居中显示
    \includegraphics[width=0.6\textwidth]{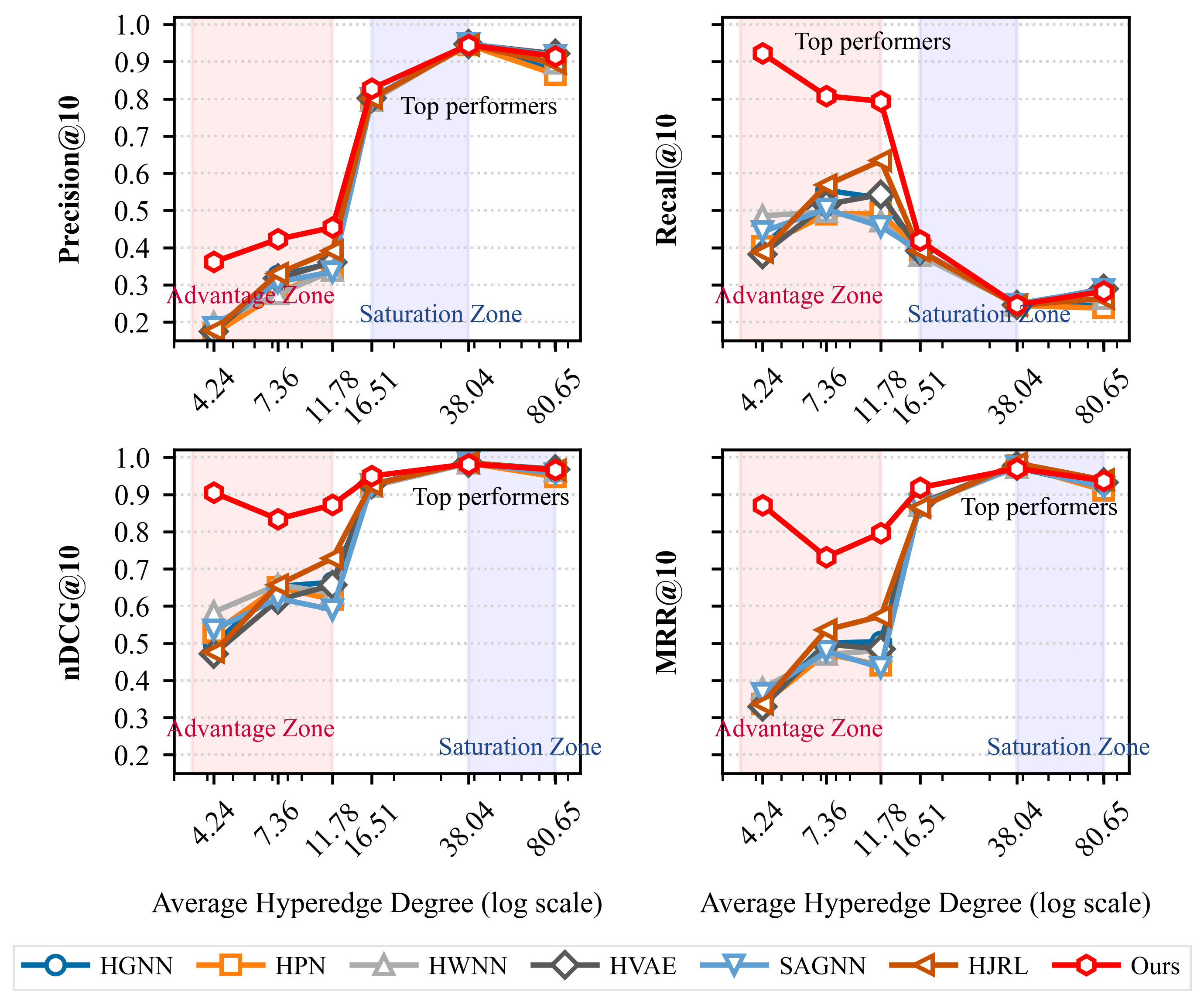} % 图片文件名(无需扩展名)
    \caption{nDCG and MRR versus hypergraph complexity} % 图片标题
    \label{complexity} % 唯一标识符(建议以fig:开头)
\end{figure*}

Figure \ref{complexity} shows that the model's recommendation accuracy improves with the structural complexity of the hypergraph, measured by the average hyperedge degree. Across Precision, NDCG, and MRR, our model gains significantly as the average hyperedge degree increases. This trend confirms that richer, more interconnected hypergraph structures enhance the model's learning capacity, resulting in higher recommendation precision.

More notably, our model consistently outperforms all baselines. This advantage is particularly evident in regimes with lower average hyperedge degrees, typically corresponding to sparser datasets. As shown in Figure~\ref{complexity}, the performance gap between our model and the baselines is largest in these regions. These findings strongly indicate that our framework is especially effective at alleviating data sparsity and cold-start issues. The robustness under sparse conditions can be attributed to the hypergraph completion mechanism, which infers and incorporates latent high-order relations, thereby enriching the learning topology even when explicit interactions are scarce.

\subsection{Influence of hypergraph completion}
\begin{figure*}[htb] % h=当前位置, t=页顶, b=页底, p=单独页
    \centering % 居中显示
    \includegraphics[width=0.8\textwidth]{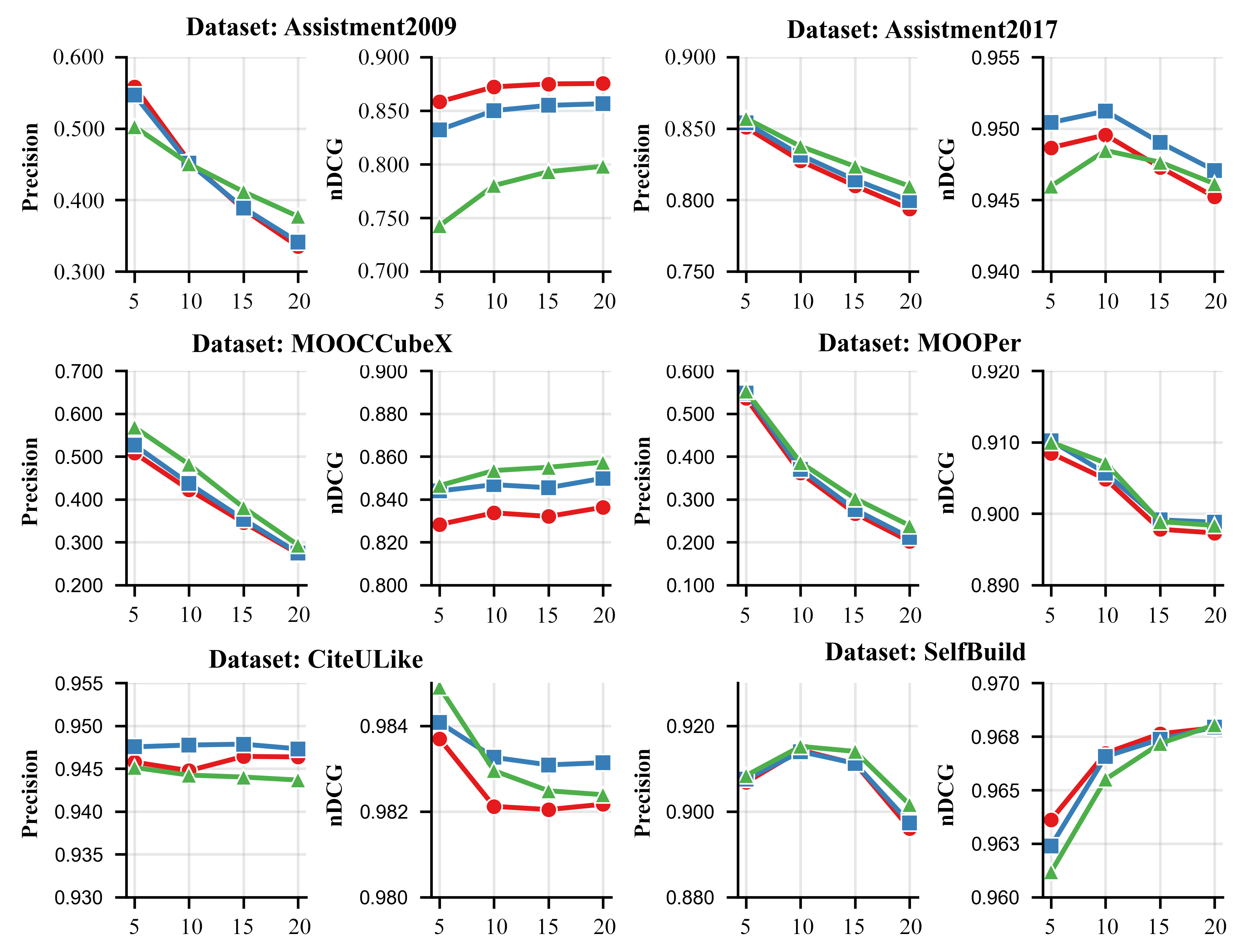} % 图片文件名(无需扩展名)
    \caption{Precision and nDCG versus $K$ with different hypergraph completion ratio} % 图片标题
    \label{completion} % 唯一标识符(建议以fig:开头)
\end{figure*}

Figure \ref{completion} illustrates the positive impact of hyperedge completion on recommendation precision. Across the majority of datasets, including Assistment2009, Assistment2017, MOOCCubeX, MOOPer, and SelfBuild, our model with hyperedge completion consistently outperforms its counterpart without completion. Precision increases steadily with a higher completion ratio, indicating that augmenting the hypergraph with inferred similarities directly strengthens the model’s capacity to retrieve relevant items.

In contrast, the impact on ranking quality, measured by nDCG, demonstrates dataset-specific heterogeneity. A positive correlation is observed on the MOOCCubeX and MOOPer datasets, where a higher completion ratio leads to better ranking performance. On the other hand, on the Assistment2009 and SelfBuild datasets, the correlation is inversely correlated. This divergence indicates that while enriching the hypergraph reliably improves precision, its effect on optimizing their relative order is not universally guaranteed. The added hyperedges may modify local structural properties in ways that do not consistently align with an optimal ranking, particularly in certain data environments.

\section{Prototype system and user study}
\label{ps}

\subsection{Context and objective}
Our proposed recommendation framework has great potential for a variety of educational applications. To validate the practical utility and demonstrate a real-world deployment scenario, we developed a fully functional prototype system designed for academic resource recommendations. In academic communities, a substantial number of researchers, such as postgraduate students, are confronted with an ever-increasing volume of scientific literature. The significant time and effort required to screen, read, and identify publications that are truly relevant to their interests can cause delays in the formation of novel research ideas and the formulation of feasible project proposals. These individuals are often overwhelmed by the sheer volume of publications, resulting in fragmented attention across multiple related topics and increased cognitive load. This reality highlights the pressing need for intelligent academic resource recommender systems that can help them efficiently navigate the literature landscape. 

Built upon the recommendation techniques proposed in this study, we implemented a prototype system and tested its effectiveness in authentic educational contexts. This system operationalizes the proposed model by providing personalized article suggestions to postgraduate students. We then conducted a controlled user study to quantitatively and qualitatively evaluate the system's usefulness and perceived recommendation quality. The goal of the system was to streamline literature discovery and help postgraduate students concentrate on high-value content, thereby boosting overall scholarly productivity.

\subsection{Prototype system design}
In this prototype system, postgraduate students browsing academic papers are referred as users, article keywords as items, and research areas as categories. The academic articles were gathered from eight high-impact Chinese core journals in the field of educational technology, all of which were published within the two years preceding the data cutoff date of September 11, 2025. A total of 1,493 articles were obtained, each with information such as the title, authors, abstract, and keywords. Based on the classification of educational technology research areas adopted by the university, these papers were categorized into three directories: Educational Technology Theory and Practice(1,083 papers), Learning Science and Technology(182 papers), and Intelligent Educational Environments and Resources(228 papers).

\begin{figure}[htbp] % h=当前位置, t=页顶, b=页底, p=单独页
    \centering % 居中显示
    \includegraphics[width=0.48\textwidth]{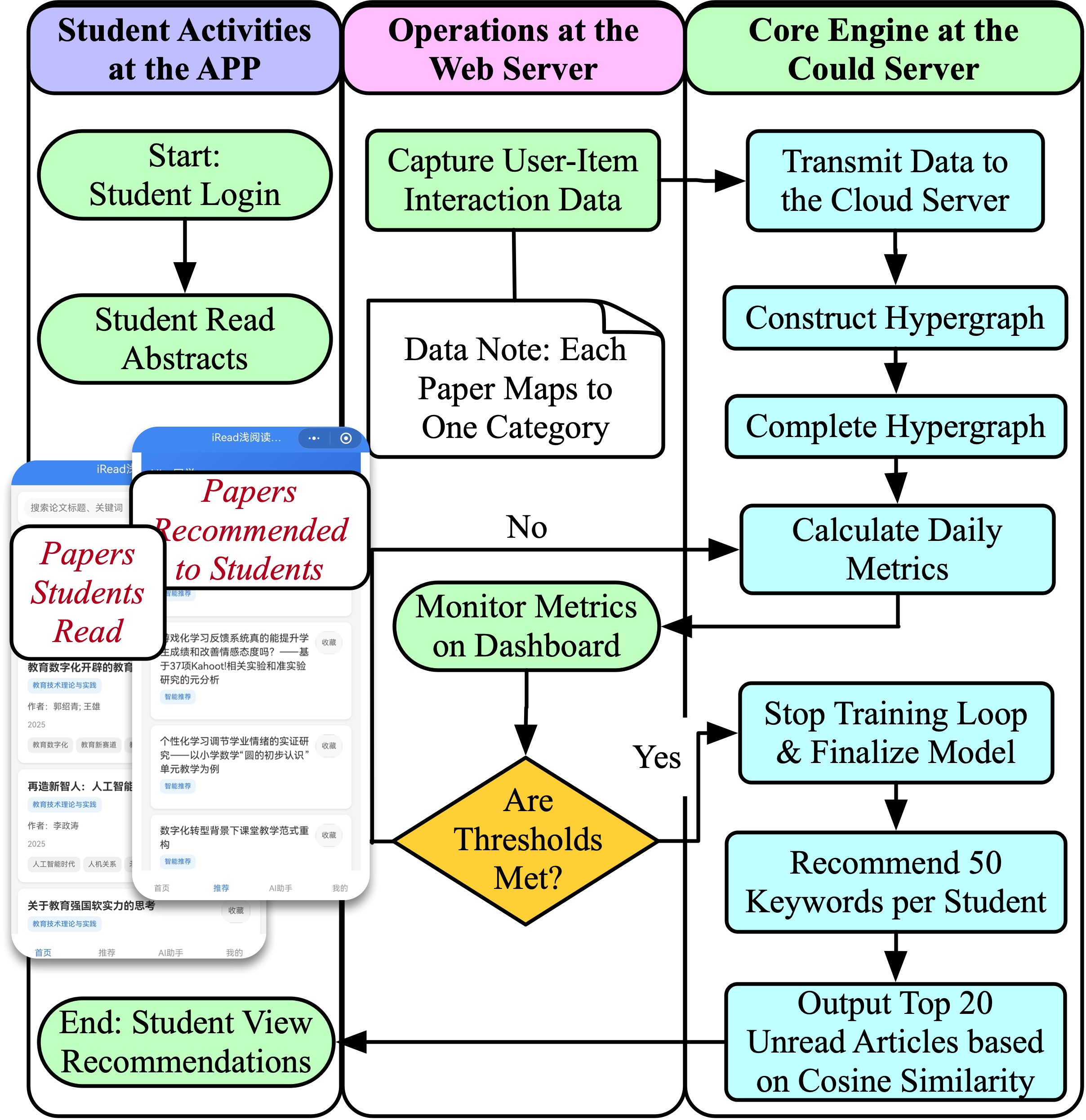} % 图片文件名(无需扩展名)
    \caption{Interaction workflow of the system} % 图片标题
    \label{interaction} % 唯一标识符(建议以fig:开头)
\end{figure}

The system architecture​ comprises a cloud server, a web server, and a mobile application client (APP). They were collectively designed to deliver full article recommendation service. The cloud server hosts the proposed recommendation model, the web server handles article and user management as well as recommendation process monitoring, and the APP serves postgraduate students, allowing them to browse articles via mobile devices. The interaction workflow​ is shown in Figure \ref{interaction}. Students freely browse abstracts of interest, and the web server records the corresponding keyword lists as interaction items. Each paper is associated with up to three research field categories. These data are transmitted to the cloud server, which performs heterogeneous hypergraph construction, hypergraph completion, and item recommendation. The web server monitors daily performance metrics, including Precision, Recall, MRR, and nDCG, until predefined thresholds are reached. Finally, the system generates 50 recommended keywords per student. These keywords are matched via Cosine similarity computation against the keywords of unread articles, producing a ranked list of the top 20 articles as the final recommendation set.

\subsection{User study}
\subsubsection{Participants}
To evaluate the practical utility and user perceptions of the prototype system, we carried out a structured user study combining quantitative surveys and qualitative interviews. A total of 32 postgraduate students majoring in Educational Technology from the same university were recruited for the user study. They received a standardized training session on the prototype system to ensure technical proficiency. After three weeks of use, when the performance metrics monitored on the web server dashboard reached predefined criteria (Precision = 0.539, Recall = 0.903, MRR = 0.780, and nDCG = 0.845), 23 participants remained actively engaged in the system. Given the observed attrition trend and the satisfactory performance achieved, the system proceeded to generate a personalized list of 20 recommended articles for each remaining user. Immediately after reviewing their recommendations, participants were asked to complete a post-study questionnaire. Subsequently, we conducted semi-structured interviews with five randomly selected participants to gather in-depth insights into their authentic experiences and perceptions of the prototype system.

\subsubsection{Instruments}
The survey questionnaire was adapted from the well-established  Recommender Systems' Quality of User Experience model (ResQue)~\cite{ain2025designing}, which integrates constructs from the Technology Acceptance Model (TAM)~\cite{pu2011user} and the Software Usability Measurement Inventory (SUMI)~\cite{kiraskowski1993sumi}. The final questionnaire comprises 15 items covering four key dimensions:

\begin{itemize}
    \item \textbf{Recommendation Quality}. It assesses the informational value of the recommended academic resources, measured in terms of perceived recommendation accuracy and familiarity.
    \item \textbf{User Beliefs}. It captures the degree to which the system is believed to support academic research, including perceived usefulness and perceived ease of use.
    \item \textbf{User Attitude}. It reflects the user's overall affective response after he/she interacts with the system, including confidence in the system and overall satisfaction.
    \item \textbf{Behavioral Intention}. It gauges users' willingness to use the system in their future research and to recommend it to peers.
\end{itemize}

All items were rated on a standard 5-point Likert scale (1 = "Strongly Disagree'' to 5 = "Strongly Agree''). The questionnaire demonstrated good reliability, with an overall Cronbach's $\alpha$ coefficient of 0.902 and all sub-dimensions exceeding the threshold of 0.7. The Kaiser-Meyer-Olkin (KMO) measure of sampling adequacy was 0.695, indicating acceptable validity for user study.

Following the quantitative survey, semi-structured interviews were conducted to collect in-depth qualitative feedback. The interview protocol focused on three core dimensions: (1) perceived effectiveness of the recommender system, (2) holistic evaluation of the user experience, and (3) open-ended improvement suggestions. A sample interview question was \emph{"To what extent do you believe the recommended academic articles align with your research interests and improve your learning efficiency? Please explain your reasoning."} This question aimed to examine the alignment between system outputs and user expectations, helping to identify potential discrepancies between self-reported behavior and actual usage patterns.

\subsubsection{Results}
Quantitative survey results demonstrate positive user evaluations across key constructs. High ratings for perceived recommendation accuracy (M = 4.72/5, SD = 0.73) and familiarity (M = 4.14/5, SD = 0.71) indicate that the system effectively aligns suggestions with users’ research interests and existing knowledge, providing a reliable basis for sustained engagement. Moreover, scores for perceived ease of use (means between 4.36/5 and 4.64/5) and perceived usefulness (means between 4.09/5 and 4.18/5) reflect solid user perceptions of the system’s usability and practical value. In terms of user attitude, participants reported considerable trust in the system (M = 4.27/5, SD = 0.63) and overall satisfaction (M = 4.18/5, SD = 0.50). Notably, the high score on the intention-to-recommend item (M = 4.23/5,SD = 0.61) suggests a strong willingness to share the system with peers. 

Qualitative analysis of interview data further revealed strong endorsement of both perceived relevance and personalization of recommendations. For instance, Student A remarked, \emph{"The recommended articles are strongly related to my reading history and align well with my interests, which significantly improves the efficiency of literature discovery."} This suggests that the system successfully aligns recommendations with users’ established reading preferences and enhances information acquisition efficiency. Additionally, respondents emphasized that the recommendations were not mere repetitions of already-read articles, but provided substantively consistent and thematically coherent literature. As Student B explained, \emph{"Most of the recommended articles fall within my ongoing research focus, offering many references of the same type and compensating for the inadequacy of my own searches."} This points to the system’s capacity to identify complementary articles within a consistent thematic scope, effectively addressing gaps in users’ current knowledge.

In summary, the above empirical findings validate the effectiveness of the proposed model and its implemented prototype system. Both quantitative results and qualitative feedback collectively confirm that the system provides accurate, relevant, and novel​ personalized recommendations that align closely with users' established research interests and knowledge domains. The high user trust, satisfaction, and intention to recommend further indicate strong user acceptance and the system's practical value for real-world academic research support.

\section{Conclusion}
\label{cp}
To address the challenges of information overload and knowledge disorientation in the knowledge era, this paper proposes a multi-view fused heterogeneous hypergraph recommendation framework with dynamic behavior profiling. The framework can effectively capture complex, high-order relationships in real-world educational scenarios while accounting for the temporal dynamics​ of user-item interactions. Through comprehensive experiments and the deployment of a prototype system, the framework has demonstrated significant improvements in recommendation accuracy, robustness, and user satisfaction, confirming its feasibility and practical relevance in authentic educational contexts. The core strength of this work lies in two aspects. First, it enhances the structural expressiveness​ of the original hypergraph through behavior profile-based dynamic modeling, leading to more powerful representations of high-order relations. Second, it achieves principled and adaptive integration of complementary information at the representation level via the multi-view fusion mechanism. This work underscores that explicitly modeling the heterogeneity, dynamics, and high-order associations​ inherent in educational data is crucial for building next-generation educational recommender systems that are adaptive and cognitively aware, which holds the potential to advance personalized education towards more intelligent and learner-centric​ paradigms. A main limitation of this study is the manual selection of the hyperedge completion rate, which may not be optimal across different datasets. Determining the appropriate granularity of hypergraph evolution automatically​ remains an open challenge that requires balancing prior knowledge, data characteristics, and computational efficiency. Future work will focus on the global optimization of structural evolution parameters​ and enhancing the educational interpretability of the learned multi-view representations.

%% Use \subsection commands to start a subsection.
%% Use \subsubsection, \paragraph, \subparagraph commands to 
%% start 3rd, 4th and 5th level sections.
%% Refer following link for more details.
%% https://en.wikibooks.org/wiki/LaTeX/Document_Structure#Sectioning_commands

%% Refer following link for more details.
%% https://en.wikibooks.org/wiki/LaTeX/Mathematics
%% https://en.wikibooks.org/wiki/LaTeX/Advanced_Mathematics

%% Use a table environment to create tables.
%% Refer following link for more details.
%% https://en.wikibooks.org/wiki/LaTeX/Tables

%% Use figure environment to create figures
%% Refer following link for more details.
%% https://en.wikibooks.org/wiki/LaTeX/Floats,_Figures_and_Captions
\section*{Acknowledgements}
The current study was supported by National Natural Science Foundation of China under Grant (No. 62277043).

\bibliographystyle{elsarticle-num} 
\bibliography{Ref}



\end{document}